\documentclass[11pt,a4paper,oneside]{article}
\usepackage{mymacro}
\usepackage{amsmath, amssymb}
\usepackage{geometry}
\usepackage{xcolor}
\usepackage{tikz}
\usepackage{genyoungtabtikz}
\def\CO{\mathcal{O}}
\def\cO{\mathcal{O}}

\newcommand{\SL}{\mathrm{SL}}

\newcommand{\cN}{\mathcal{N}}

\newcommand{\fa}{\mathfrak{a}}
\newcommand{\dfa}{\dot{\mathfrak{a}}}
\newcommand{\fb}{\mathfrak{b}}
\newcommand{\dfb}{\dot{\mathfrak{b}}}
\newcommand{\ula}{\underline{\lambda}}

\allowdisplaybreaks

\title{\textbf{Non-Renormalizability Properties of Quarter-BPS Correlators in $\mathcal{N}$=4 SYM}}

\author{Agnese Bissi$^{a,b,c}$,}
\author{Giulia Fardelli$^{d,e}$}
\author{and Hector Puerta-Ramisa$^f$}

\affiliation{$^{a}$ICTP, International Centre for Theoretical Physics, \\
				Strada Costiera 11, 34151, Trieste, Italy}
\affiliation{$^{b}$INFN, Sezione di Trieste, Via Valerio 2, 34127 Trieste, Italy,}
\affiliation{$^{c}$Department of Physics and Astronomy, Uppsala University,\\
Box 516, SE-751 20 Uppsala, Sweden}
\affiliation{$^{d}$Department of Physics, Boston University, Boston, MA 02215, USA}
\affiliation{$^{e}$Fields and Strings Laboratory, Institute of Physics, École Polytechnique Fédéral de Lausanne (EPFL),
Route de la Sorge, CH-1015 Lausanne, Switzerland}
\affiliation{$^{f}$Department of Mathematical Sciences, Durham University, \\
Durham, DH1 3LE, United Kingdom}

\makeatletter
\renewcommand{\@email}[1]{#1}
\makeatother

\emailAdd{\href{mailto:abissi@ictp.it}{\tt abissi@ictp.it},
            \href{mailto:giulia.fardelli@epfl.ch}{\tt giulia.fardelli@epfl.ch},
             \href{mailto:hector.puerta-ramisa@durham.ac.uk}{\tt hector.puerta-ramisa@durham.ac.uk}
           
}

\abstract{We study extremal and next-to-extremal four-point correlators involving general quarter-BPS operators in $\mathcal{N}$=4 Super Yang-Mills. Using a combination of $R$-symmetry considerations, operator product expansion content, and analytic superspace techniques, we show that such correlators get contributions only from operators belonging to short and semi-short multiplets. This result together with the non-renormalizability of the relevant three-point functions guarantees that all such four-point correlators are protected from quantum corrections.
}

\begin{document}

\maketitle

\section{Introduction}

Four-dimensional \(\mathcal{N}=4\) Super Yang--Mills (SYM) theory occupies a privileged
position in the landscape of quantum field theories.  Its exact conformal
invariance for any value of the coupling, its maximal superconformal symmetry,
and its duality with type IIB string theory on
\(\mathrm{AdS}_5\times \text{S}^5\)  make it a unique meeting
point for integrability, supersymmetric localization, the analytic conformal
bootstrap, and holography.  A persistent theme in this programme is the
interplay between protected and dynamical observables: the former are
controlled by supersymmetry and often computable exactly, while the latter
probe the coupling-dependent regime and, ultimately, the gravitational
physics of the holographic dual.
 
The simplest protected observables are built from half-BPS primaries
\(\mathcal{O}_{p}\), scalar operators of protected dimension \(\Delta=p\)
transforming in the \([0,p,0]\) representation of the \(\mathrm{SU}(4)_{R}\) $R$-symmetry group.
Two- and three-point functions of half-BPS operators are independent of the
gauge coupling and given by their free-field values. 
Four-point functions are, by contrast, generically dynamical: the superconformal
Ward identities organise them into a free piece together with a reduced
correlator \(\mathcal{H}(u,v;\sigma,\tau)\) that encodes the exchange of
long, unprotected multiplets \cite{Dolan:2004iy,Beem:2016wfs}.  A distinguished family of exceptions arises
for the extremal ($k=0$) and next-to-extremal ($k=1$) four-point functions of schematic form
\begin{align}
\langle \mathcal{O}_{p_{1}}\mathcal{O}_{p_{2}}\mathcal{O}_{p_{3}}\mathcal{O}_{p_{4}}\rangle\,,
\qquad
p_{1}=p_{2}+p_{3}+p_{4}-2k\,,\qquad k\in\{0,1\}\,,
\label{eq:halfBPSextremal}
\end{align}
which were shown long ago to be entirely protected from quantum corrections \cite{DHoker:1999jke,DHoker:2000xhf,Bianchi:1999ie,Erdmenger:1999pz}.\footnote{Related results for $n$-point functions or different amounts of supersymmetry can be found in \cite{Eden:1999kw,Eden:2000gg,Eden:2001ec,Arutyunov:2001qw}.}
 
A natural question is whether similar considerations apply to operators with lower amounts of supersymmetry, such as quarter-BPS operators.
Such operators are scalars transforming in \([q,p,q]\) representations of
\(\mathrm{SU}(4)_{R}\) and have protected dimension \(\Delta=p+2q\).  They are
intrinsically multi-trace at large \(N\), and similarly to the half-BPS case, their three-point couplings in general
are protected  \cite{ Heslop:2001gp,Heslop:2003xu,Basu:2004nt}.  From the holographic perspective they are
dual to multi-particle excitations on \(\mathrm{AdS}_{5}\times \text{S}^{5}\), and
they provide one of the cleanest probes of non-planar physics and multi-trace mixing. Correlation functions involving quarter-BPS operators were studied in early works~\cite{Ryzhov:2001bp,DHoker:2001jzy,DHoker:2003csh}, with renewed interest in this direction in more recent studies~\cite{Bissi:2021hjk, Ganchev:2025dzn}.

In particular they play a central role in understanding how to construct corrections to the four-point functions of half-BPS operators in the limit of large $N$ \cite{Bissi:2024tqf, Bissi:2021hjk, Bissi:2020woe, Bissi:2020wtv}.  Pinning down the protected
sector of quarter-BPS correlators is therefore a prerequisite both for the
analytic bootstrap of multi-trace four-point functions and for any attempt
to use these correlators as a window onto quantum-gravitational corrections.
 
This motivates the question we address in the present paper:
which four-point functions involving quarter-BPS operators inherit a
non-renormalization theorem of the type known for half-BPS correlators?

Non-renormalization theorems are far more than a catalogue of coincidences: they form the scaffolding on which much of the solvability of $\mathcal{N}=4$ SYM rests. The statement that two- and
three-point functions of half-BPS operators are coupling-independent 
fixes the leading OPE data of the theory, and the extremal and next-to-extremal
non-renormalization theorems for higher-point functions
extend this rigidity to an infinite family of correlators. These protected quantities are exactly
the boundary data that render the modern analytic bootstrap of $\mathcal{N}=4$ SYM tractable: they
supply the exactly known input against which the coupling-dependent, dynamical part of a
correlator is solved. A systematic description
of the protected sector is therefore a precondition for extracting the dynamical sector
reliably.

A central object in the analytic bootstrap is the superconformal block:
the kinematic function that captures the contribution of an entire
exchanged supermultiplet, a superprimary together with all of its
superdescendants, to a four-point function, with all dynamical
information confined to the spectrum of exchanged operators and their OPE 
data \cite{Nirschl:2004pa}. These blocks are fixed by the superconformal algebra, being eigenfunctions of the superconformal Casimir operator, and they are
obtained precisely by solving the superconformal Ward identities. For
quarter-BPS operators we can carry out this step explicitly in analytic
superspace, which is what makes their blocks accessible: the solution not
only identifies which multiplets are exchanged in a given channel,
but produces the independent three-point tensor structures associated with
each exchange. The latter is essential, because quarter-BPS primaries
carry non-trivial $\SU(4)_R$ structure and are therefore exchanged through
several structures at once. Extremal and
next-to-extremal correlators are the ideal setting in which to pin this down:
their content is fixed entirely by representation theory and generically kinematical statements, with only short
and semi-short multiplets exchanged.

The paper is organised as follows.

In section \ref{sec:Rsym}, we show the known result that extremal and next-to-extremal four-point functions of half-BPS operators do not exchange long superconformal multiplets by intersecting the $\SU(4)_R$ selection rules for the long sector of both OPEs. 
In the half-BPS case, this simple representation theory argument is enough to rule out all long superconformal multiplets, since the half-BPS OPE preserves the $R$-symmetry selection rule at the level of superprimaries (i.e., the superprimary three-point function does not vanish). This result implies that the correlator is protected because three-point functions involving two half-BPS operators and a third short or semi-short multiplet are known to be protected. 

The same argument can be applied to correlators involving two quarter-BPS operators and two half-BPS ones, yielding the same result: no long operators sit at the intersection. However, this does not imply the full correlator is protected, since there is a sector of exchanged superconformal multiplets whose superprimaries do not preserve the $R$-symmetry selection rule, so is not captured by the $R$-symmetry argument. Still, this partial proof is highly suggestive as generically both $R$-symmetry preserving and violating sectors are related, which is most easily  seen in superspace formalisms.

To complete the above exercise, and to handle correlators with more quarter-BPS insertions, we turn to analytic superspace \cite{Howe:1995md}. This formalism has the key advantage that all representations are automatically irreducible \cite{Heslop:2001zm}. This means that shortening conditions are satisfied automatically and thus do not have to be imposed as additional differential Ward identities. Instead, correlation functions are required to be pole-free in certain ``internal'' variables $y$ associated with the $R$-symmetry \cite{Heslop:2001gp,Heslop:2003xu,Heslop:2025zeq}. This requirement, known as harmonic analyticity, replaces the difficult differential computations involved in imposing shortening conditions by simple algebraic constraints. We review these ideas and introduce our main method in section \ref{sec:superspace}.

Then, in section \ref{sec:generalqbps}, we use these methods to 
derive the intersection of the two superconformal OPEs in four-point functions of general quarter-BPS operators,
\begin{equation}
    \langle \cO_{[q_1,p_1,q_1]} \cO_{[q_2,p_2,q_2]}\cO_{[q_3,p_3,q_3]} \cO_{[q_4,p_4,q_4]}\rangle,
\end{equation}
directly in analytic superspace. We prove in the extremal case, that is, when
\begin{equation}
    p_1+2q_1= p_2+2q_2+p_3+2q_3+p_4+2q_4,
\end{equation}
that only short (half- and quarter-BPS) multiplets are exchanged, and in fact the full spectrum can be obtained from intersecting the $R$-symmetry selection rule from both OPEs. 
This generalises the result of \cite{Heslop:2001gp} involving one quarter-BPS operator in the $[1,p,1]$ and three half-BPS operators. 
The
next-to-extremal case, 
\begin{equation}
    p_1+2q_1= p_2+2q_2+p_3+2q_3+p_4+2q_4-2,
\end{equation}
is far less constrained but the exchanged spectrum is shown to consist of short and semi-short multiplets only. In all cases, not only do we rule out long superconformal multiplets, we also classify the specific exchanges, and construct the corresponding three-point tensor structures describing each OPE.

The promotion to a full non-renormalization theorem for the four-point functions is discussed in section \ref{sec:nonrenorm}.

Finally, in appendix~\ref{app:explicitO02O2O2O2}, we present an explicit computation of a next-to-extremal correlator involving three half-BPS $\mathcal{O}_2$ operators and a quarter-BPS operator transforming in the $[2,0,2]$ representation. We analyse this example both in superspace and through an explicit free-theory computation. We illustrate how the different exchanges contribute to the four-point function, by connecting analytic superspace predictions with the results from the chiral algebra~\cite{Beem:2013sza,Beem:2016wfs},  and provide some insight into the possible construction of superblocks.

\section{$R$-symmetry considerations}\label{sec:Rsym}
Extremal and next-to-extremal half-BPS four-point functions can be shown to not acquire quantum corrections by simple group-theoretic arguments on the $R$-symmetry group $\SU(4)_R$. In this section, we introduce this method and apply it to correlators with up to two quarter-BPS operators. 
\subsection{Four half-BPS operators}
We consider four-point functions made of half-BPS operators, denoted by 
$\mathcal{O}_p$. They are scalar operators, of dimension $\Delta = p$ and
transforming in the representation $[0,p,0]$ of $\SU(4)_R$. Their dimension and OPE coefficients are protected from quantum corrections, thus they only depend on $N$, the rank of the gauge group. Generically, four-point functions of  half-BPS operators depend on the coupling constant $\lambda$. However, in the case of extremal or next-to-extremal correlators, four-point functions are also protected \cite{DHoker:1999jke,DHoker:2000xhf,Bianchi:1999ie}. 
In what follows, we will analyse how to understand this phenomenon by looking at the selection rules for long operators. 

To simplify the notation we start with correlators of identical operators and we contract the SU(4)$_R$ indices with a null six-dimensional auxiliary vector $\mathrm{y}$. The four-point function by conformal invariance takes the form
\begin{equation}
    \langle \mathcal{O}_p(x_1,\mathrm{y}_1) \cdots \mathcal{O}_p(x_4,\mathrm{y}_4) \rangle
    \;=\;
    \frac{(\mathrm{y}_1 \cdot \mathrm{y}_2)^p (\mathrm{y}_3 \cdot \mathrm{y}_4)^p}{x_{12}^{2p}\, x_{34}^{2p}}
    \;\mathcal{G}(u,v;\sigma,\tau) \,,
    \label{eq:4pt-structure}
\end{equation}
where $x_{ij}^2 \equiv (x_i - x_j)^2$, and $\mathcal{G}(u,v;\sigma,\tau)$ is a function of four cross-ratios.
The two space-time (conformal) cross-ratios are
\begin{equation}
    u \;=\; z\bar{z} \;=\; \frac{x_{12}^2 x_{34}^2}{x_{13}^2 x_{24}^2}\,,
    \qquad
    v \;=\; (1-z)(1-\bar{z}) \;=\; \frac{x_{14}^2 x_{23}^2}{x_{13}^2 x_{24}^2}\,.
    \label{eq:cross-ratios}
\end{equation}
The two $R$-symmetry cross-ratios are defined analogously using the null vectors $\mathrm{y}_i$:
\begin{equation}
    \sigma \;=\; \alpha\bar{\alpha} \;=\; \frac{(\mathrm{y}_1 \cdot \mathrm{y}_2)(\mathrm{y}_3 \cdot \mathrm{y}_4)}{(\mathrm{y}_1 \cdot \mathrm{y}_3)(\mathrm{y}_2 \cdot \mathrm{y}_4)}\,,
    \qquad
    \tau \;=\; (1-\alpha)(1-\bar{\alpha}) \;=\; \frac{(\mathrm{y}_1 \cdot \mathrm{y}_4)(\mathrm{y}_2 \cdot \mathrm{y}_3)}{(\mathrm{y}_1 \cdot \mathrm{y}_3)(\mathrm{y}_2 \cdot \mathrm{y}_4)}\,.
    \label{eq:R-cross-ratios}
\end{equation}
The variables $(z, \bar{z})$ and $(\alpha, \bar\alpha)$ are the holomorphic/antiholomorphic splitting of the conformal and $R$-symmetry cross-ratios, respectively.
The four-point function \eqref{eq:4pt-structure} is further constrained by the supercharges of $\text{PSU}(2,2|4)$. The superconformal Ward identities impose that $\mathcal{G}$ satisfies a set of first-order differential equations. Their solution can be written as
\begin{equation}
    \mathcal{G}(u,v;\sigma,\tau) \;=\; \mathcal{F}_{\mathrm{free}}(u,v;\sigma,\tau) \;+\; \mathcal{H}(u,v;\sigma,\tau)\,,
    \label{eq:G-split}
\end{equation}
where $\mathcal{F}_{\mathrm{free}}$ is the free-field contribution, and $\mathcal{H}$ is a reduced correlator taking into account the contribution of long, non-protected operators. 

It is instructive to consider explicitly extremal and next-to-extremal correlators made of half-BPS
operators. In the simplest case in which three of them are identical:
\twoseqn{
&\langle \mathcal{O}_p(x_1, \mathrm{y}_1)\, \mathcal{O}_p(x_2, \mathrm{y}_2)\, \mathcal{O}_p(x_3, \mathrm{y}_3)\,
    \mathcal{O}_{3p}(x_4, \mathrm{y}_4) \rangle
    && \text{(extremal)},
}[]
{
&\langle \mathcal{O}_p(x_1, \mathrm{y}_1)\, \mathcal{O}_p(x_2, \mathrm{y}_2)\, \mathcal{O}_p(x_3, \mathrm{y}_3)\,
    \mathcal{O}_{3p-2}(x_4, \mathrm{y}_4) \rangle
    && \text{(next-to-extremal)}.
}[][4ptboth]
 
The intermediate conformal primary operators contributing to \eqref{4ptboth} transform under the representations that appear in
\begin{equation}
    [0,p,0] \otimes [0,p,0]
    \;\cap\;
    [0,p,0] \otimes [0,3p,0]=[0,2p,0],
\end{equation}
and
\begin{equation}
    [0,p,0] \otimes [0,p,0]
    \;\cap\;
    [0,p,0] \otimes [0,3p-2,0]
    \;=\;
    [0,2p,0] \oplus [0,2p-2,0] \oplus [1,2p-2,1].
\end{equation}
 
In order to understand which of these representations host long superconformal primaries, we need
to invoke superconformal Ward identities. For the half-BPS case their solution is known,
and for these cases they read
\begin{equation}
    [0,p,0] \otimes [0,p,0] \Big|_{\rm long}
    \;=\;
    \sum_{\substack{0 \leq m \leq n \leq p-2}}
    [n-m,\, 2m,\, n-m].
\end{equation}

A simple observation to make is that
\begin{equation}
    \Bigl([0,2p,0] \oplus [0,2p-2,0] \oplus [1,2p-2,1]\Bigr)
    \;\cap\;
    \sum_{\substack{0 \leq m \leq n \leq p-2}}
    [n-m,\, 2m,\, n-m]
    \;=\; \varnothing.
\end{equation}
 
For example:
\begin{itemize}
    \item $p=2$:\quad
    $\bigl([0,4,0] \oplus [0,2,0] \oplus [1,2,1]\bigr)
    \cap [0,0,0] = \varnothing$,
 
    \item $p=3$:\quad
    $\bigl([0,6,0] \oplus [0,4,0] \oplus [1,4,1]\bigr)
    \cap \bigl([0,0,0] \oplus [0,2,0] \oplus [1,2,1]\bigr) = \varnothing$.
\end{itemize}
 
This means that there is no possibility for long intermediate operators to appear in the OPE of
extremal and next-to-extremal correlators made of $\frac{1}{2}$-BPS operators, so only operators belonging to short and semi-short representations can appear in the OPE decompositions of \eqref{4ptboth}. Furthermore, since all three-point function coefficients involved (two half-BPS and one short or semi-short operator) are also protected, we conclude that extremal and next-to-extremal correlators are protected from quantum corrections: they do not acquire corrections in the coupling $\lambda$ and depend only on $N$.

This argument makes an important assumption: the superconformal OPE of half-BPS operators satisfies the $R$-symmetry selection rule. This means that the superprimary three-point functions describing the OPE are non-vanishing. For half-BPS OPEs this statement is true, since all superprimaries in the OPE are constructed from products and derivatives of the two scalar half-BPS superprimaries. This property is not satisfied by OPEs preserving less supersymmetry.

We note that multiplet recombinations are not possible, that is, short and semi-short multiplets cannot combine into a long superconformal multiplet at threshold. For this to happen, there must be a long multiplet whose superprimary has the $\SU(4)_R$ representation of a tower of semi-shorts, which is ruled out in extremal and next-to-extremal correlators as we showed here.

The same argument can be straightforwardly applied to a correlator of half-BPS operators with unequal charges $p_1,p_2,p_3,p_4$. We will instead prove this general case  in section \ref{sec:generalqbps} using analytic superspace arguments (introduced in section \ref{sec:superspace}) in a manner that generalises straightforwardly to quarter-BPS correlators.

\subsection{Two half-BPS and two quarter-BPS}
The same $R$-symmetry selection rule argument can be applied to four-point functions involving up to two quarter-BPS operators, but the result is not as strong as in the fully half-BPS case, since the quarter-BPS OPE contains superconformal multiplets whose superprimaries do not obey the $R$-symmetry selection rule. We consider an example and expand on this point below.

Quarter-BPS are, after half-BPS, the shortest superconformal multiplets in $\cN=4$ SCFTs in $4d$.
They can be labelled in terms of the quantum numbers of the  corresponding superprimaries, which are  Lorentz scalars  transforming under $R$-symmetry in the 
\begin{equation}
    [q,p,q],
\end{equation}
and with protected dimension $\Delta=2q+p$.
We thus denote them by $\cO_{[q,p,q]}$. We refer to \cite{DHoker:2003csh,Bissi:2021hjk} for further background.

Let us consider the following correlator
\begin{equation}
   \langle  \cO_{p_1} \cO_{p_2} \cO_{[q_1,0,q_1]} \cO_{[q_2,0,q_2]}\rangle.
\end{equation}
Without loss of generality we assume $q_2\ge q_1$ and $p_2\geq p_1$. In the extremal case we have,
\begin{align}
    p_1+p_2+2q_1=2q_2,
\end{align}
which requires $p_1+p_2$ to be a positive even integer. So let us define
\begin{align}
    p_2=p_1+2\delta\, , \qquad \delta \in \mathbb{N}\,.
\end{align}
In the $s$-channel OPE we know that the only long operators that can be exchanged in the $\CO_{p_1}\times \cO_{p_2}$ OPE must transform in the
\begin{align}\label{eq:HHOPE}
   \sum_{n=0}^{\text{min}(p_1-2,p_2-2)}\sum_{m=0}^n [n-m, 2m+|p_{12}|, n-m]=\sum_{n=0}^{p_1-2}\sum_{m=0}^n [n-m, 2m+2\delta, n-m]\,,
\end{align}
where we have used our assumptions. Next, we need to consider the intersection with the OPE involving quarter-BPS operators. Naively in this case, we can just consider selection based on SU(4)$_R$ symmetry and just look at the representations in the tensor product of the form $[a,b,a]$. The only ones with a non-vanishing overlap with~\eqref{eq:HHOPE} are
\begin{align}
    [q,0,q]\!\otimes\! \left[ q+\frac{p_1+p_2}{2},0, q+\frac{p_1+p_2}{2}\right] \supset \!\sum_{\beta=0}^q\!\sum_{\alpha=0}^{2(q-\beta)}\left[\alpha+\frac{p_1+p_2}{2}, 2\beta,\alpha+\frac{p_1+p_2}{2}\right]\,.
\end{align}
However to have a non-vanishing overlap with~\eqref{eq:HHOPE} we also need to impose
\begin{align}
\begin{cases}
 \beta=m+\delta\,,\\
\alpha+p_1+\delta=n-m,
\end{cases}
\quad \Longrightarrow \quad n-p_1=\alpha+\beta \, ,
\end{align}
but $n\leq p_1-2$ and $\alpha+\beta \ge 0 $, so this condition can never be satisfied! In the case of next-to-extremal similarly we find
\begin{align}
    n-(p_1-1)=\alpha+\beta\, ,
\end{align}
which once again is never satisfied.

As stated at the beginning of the section, this exercise only rules out long superconformal multiplets in the $R$-symmetry preserving sector of the OPE. This is the sector where the corresponding superprimary three-point functions are non-vanishing. Unlike in the half-BPS OPE, here the $R$-symmetry violating sector is non-trivial, because of the existence of superprimaries of the schematic form
\begin{equation}
   (Q_{\alpha} \cO_{\frac14}) \cO \in  \cO_{\frac14} \times \cO, 
\end{equation}
where $\cO_{\frac14}$ is a quarter-BPS superprimary state, $\cO$ is an arbitrary superprimary (including half-BPS) and $Q_{\alpha}$ is a supercharge. The above superprimary vanishes in a half-BPS OPE by the half-BPS shortening condition.

The existence of $R$-symmetry preserving and violating sectors is an artifact of working in position space at the level of conformal multiplets. In superspace, there is no such distinction. Therefore, the above argument provides good evidence for the stronger result that no long superconformal multiplets are exchanged. We prove this in analytic superspace below.

\section{Superconformal Ward identities and harmonic analyticity}\label{sec:superspace}

We review the key background on analytic superspace \cite{Howe:1995md,Heslop:2001zm,Heslop:2003xu,Doobary:2015gia,Aprile:2021pwd,Heslop:2022xgp,Hansen:2025lig} that is needed for setting up and solving superconformal Ward identities for correlators of quarter-BPS operators. 
This includes using harmonic analyticity in a manner similar to \cite{Heslop:2001gp,Heslop:2003xu,Heslop:2025zeq} though applied to correlators of general quarter-BPS operators.

\Yboxdim{6pt}
\subsection{Representations in analytic superspace}
Analytic superspace is one of the most powerful formalisms for dealing with short (super) conformal representations. This is because such multiplets are automatically irreducible, as proved in \cite{Heslop:2001zm}. This means that shortening conditions, typically enforced via differential constraints, need not be applied. Instead, all superconformal Ward identities are solved by simple algebraic constraints on the number of (super) indices, homogeneity and harmonic analyticity, to be defined.

In four-dimensional $\cN=4$ SCFT, analytic superspace is obtained by supplementing Minkowski superspace with internal coordinates $y^{aa'}$ on the $R$-symmetry flag manifold $$\SU(4)/\mathrm{S}(\mathrm{U}(2)\times \mathrm{U}(2)).$$ Its local coordinates are given by $(2|2)\times(2|2)$ matrices of the form
\begin{equation}\label{eq:superspacecoord}
    X^{\fa \dfa} = \begin{pmatrix}
        x^{\alpha \dot{\alpha}} & \rho^{\alpha a'}\\
        \bar{\rho}^{a \dot{\alpha}} & y^{aa'}
    \end{pmatrix},
\end{equation}
where $\alpha, \dot{\alpha}$ are Lorentz spinor indices, $a,a'$ label the fundamentals of the two internal $\SU(2)$ factors splitting the $R$-symmetry $\SU(4)_R$, and $\rho,\bar{\rho}$ are fermionic coordinates generating the supermultiplets. The latter make up half of the sixteen odd coordinates of the full $\cN=4$ supersymmetry.

Superconformal primaries and their descendants transform covariantly under the Levi subgroup
\begin{equation}
    \SL(2|2)\times \SL(2|2) \times \mathbb{C}^*,
\end{equation}
that is, they are labelled by two $\SL(2|2)$ Young diagrams $\ula_L$ and $\ula_R$, and an integer charge $\gamma$.
The Young diagrams should be thought of as containing $(2|2)$-dimensional indices $\fa, \fb, \ldots$ in $\ula_L$, $\dfa,\dfb, \ldots$ in $\ula_R$, symmetrised accordingly. These are a supersymmetric generalisation of Weyl spinor indices, which are symmetrised by definition since antisymmetrising them in $\SL(2)$ returns the determinant representation. $\SL(2|2)$ Young diagrams fit into the following thick hook containing the window diagram $\Yvcentermath1 \yng(2,2)$\,, 
\begin{equation}
    \ula_L,\, \ula_R \ \  \subset \ \ \begin{tikzpicture}[scale=0.65,baseline=(current bounding box.center)]
        \draw[thick] (1,0)--(4,0);
        \draw[thick,dotted] (4,0)--(5,0);
        \draw[thick] (1,-1)--(4,-1);
        \draw[thick,dotted] (4,-1)--(5,-1);
        \draw[thick] (0,0)rectangle(1,-1);
        \draw[thick] (0,-0.5)--(1,-0.5);
        \draw[thick] (0.5,0)--(0.5,-1);
        \draw[thick] (1,-1)--(1,-3);
        \draw[thick,dotted] (1,-3)--(1,-3.7);
        \draw[thick] (0,-1)--(0,-3);
        \draw[thick,dotted] (0,-3)--(0,-3.7);
    \end{tikzpicture}\ .
\end{equation}
We thus write a full superconformal multiplet in analytic superspace as follows,
\begin{equation}
    \cO^\gamma_{\ula_L,\ula_R}.
\end{equation}
This is an alternative and more compact labelling compared to the standard spin $[s,\bar{s}]$, dimension $\Delta$, $R$-symmetry $[R_1,R_2,R_3]$, and shortening condition. The translation for the case $\ula_L=\ula_R$, known as symmetric representations, is 
\begin{equation}\label{eq:reptranslation}
 \ula_L=\ula_R:=  \ula= [1^{\mu_1}, 1^{\mu_2}, \lambda_1,\lambda_2]\ = \ \begin{tikzpicture}[scale=0.7, x=1.3cm, y=1.3cm, baseline=(current bounding box.center)]
    \draw [thick] (-0.5,0) rectangle (0,-2);
    \node at (-0.25,-1.75) {$\mu_2$};
    \draw [thick] (-1,0) rectangle (-0.5,-3);
    \node at (-0.75,-2.75) {$\mu_1$};
    \draw [<->] (0,-2.05) -- (0,-3) node[midway,right] {$R_1=R_3$};
    \draw [dashed] (-0.5,-3) -- (0,-3);
        \draw [thick] (0,0) rectangle (3,-0.5);
        \draw [dashed] (3, -0.5) -- (3, -1);
        \draw [<->] (2.05,-1) -- (3,-1) node[midway,below] {$s=\bar{s}$};
        \draw [thick] (0,-0.5) rectangle (2,-1);
        \node at (2.75,-0.25) {$\lambda_1$};
        \node at (1.75,-0.75) {$\lambda_2$};
    \end{tikzpicture} \, ,
\end{equation}
with
\begin{equation}
 \Delta=\gamma +\sum_i \lambda_i, \qquad R_1= R_3 =\mu_1-\mu_2, \qquad R_2=\gamma-2\mu_1.
\end{equation}
The generalisation to non-symmetric representations is straightforward and will lead to representations with $s\neq \bar{s}$ and $R_1\neq R_3$ \cite{Hansen:2025lig}. The main takeaway is that columns encode $R$-symmetry and rows encode the spin and dimension. 

Shortening conditions can also be directly read from the Young diagrams without first needing to compare with unitarity bounds. Indeed, a representation is long if and only if it contains the window Young diagram,
\begin{equation}\label{eq:longYD}\Yvcentermath1\Yboxdim{8pt}
    \ula_{\text{long}}\  \supseteq \  \yng(2,2)\, .
\end{equation}
Such representations are known as \textit{typical} in the representation theory literature.
Representations satisfying shortening conditions are known as \textit{atypical} and are classified as follows,
\begin{equation}\label{eq:shortdiagram}
    \begin{aligned}
        &\text{Half-BPS:}&\qquad \ula&=\bullet \, ,\\
        &\text{Quarter-BPS:} & \ula&=[1^\mu] = \begin{tikzpicture}[scale=0.65, baseline=(current bounding box.center)]
        \draw[thick] (0,0) rectangle (0.5,2.2);
        \draw[<->] (0.75, 0) -- (0.75, 2.2) node[midway,right] {$\mu$};
    \end{tikzpicture}\, ,\\
        &\text{Semi-short:} & \ula&=[1^\mu,\lambda]=\begin{tikzpicture}[scale=0.65, baseline=(current bounding box.center)]
        \draw[thick] (0,0) rectangle (0.5,2.2);
         \draw[thick] (0.5,2.2) rectangle (3,1.7);
        \draw[<->] (-0.25, 0) -- (-0.25, 2.2) node[midway,left] {$\mu$};
        \draw[<->] (0.55, 1.45) -- (3, 1.45) node[midway,below] {$\lambda$};
    \end{tikzpicture}\ ,
    \end{aligned}
\end{equation}
where $\ula=\bullet $ denotes the trivial Young diagram, that is, half-BPS operators are scalars in analytic superspace.

Half- and quarter-BPS representations are isolated, meaning there are no nearby unitary representations.
Moreover, all half-BPS and a subsector of quarter-BPS operators ($\mu=1$ in \eqref{eq:shortdiagram}) are absolutely protected, so they do not acquire anomalous dimensions under exactly marginal deformations of the theory. 
General $\mu>1$ quarter-BPS operators do not directly satisfy this property, because they can participate in recombination with long superconformal multiplets \cite{Cordova:2016emh}. Equivalently, given a basis of free states satisfying the quarter-BPS superprimary condition relating dimension and $R$-symmetry, turning on interactions will mix these and only certain linear combinations will be truly protected. The states that do acquire anomalous dimensions can always be written as superdescendants in a long superconformal multiplet. See \cite{DHoker:2003csh} for a systematic construction of quarter-BPS superprimaries with protected dimension. The arguments we will use throughout the paper are from superconformal representation theory, and thus quarter-BPS states are assumed to be the correct superprimaries. 

Semi-short multiplets on the other hand are at threshold, and thus do acquire anomalous dimension by mixing with long superconformal multiplets with the same $R$-symmetry representation. In the context of a four-point function, this mixing can only occur if both semi-short and long representations are exchanged.

Non-symmetric representations, e.g. $\frac18$ or $\frac1{16}$-BPS, mix shortening conditions but must always contain the same number of boxes on each side $|\ula_L|=|\ula_R|$. This additional constraint is due to the additional projective constraint on the $d=\cN=4$ superconformal group $\mathrm{PSU}(2,2|4)$, rather than $\SU(2,2|4)$.

When writing superconformal multiplets, particularly when tabulating all exchanges in a four-point function, we will use the analytic superspace notation $\cO^\gamma_{\ula_L,\ula_R}$ introduced in this section, as well as the notation of \cite{Cordova:2016emh} for the reader's convenience. 
Superconformal multiplets are written in the latter as
\begin{equation}\label{eq:multipletnotation}
    {\cal S} {\cal\bar{S}} [s,\bar{s}]_{\Delta}^{[R_1,R_2,R_3]},
\end{equation}
where $s,\bar{s},\Delta,R_i$ describe the superprimary in the usual manner, and $S$, $\bar{S}$ denote the $Q$ and $\bar{Q}$ shortening conditions. Where $ {\cal S}=B$ means isolated (half or quarter-BPS), $ {\cal S}=A$ means at threshold and $ {\cal S}=L$ means long. When shortening conditions are equal on both sides we will only write one letter. Isolated multiplets will often be more simply denoted by $\cO_{p}$ for half-BPS and $\cO_{[q,p,q]}$ for quarter-BPS as defined in section \ref{sec:Rsym}.

\subsection{Superconformal Ward identities}\label{subsec:Wardids}

\subsubsection{Setup}
Solving superconformal Ward identities in analytic superspace is the natural supersymmetric analogue of solving conformal Ward identities in Minkowski space using Weyl spinors.  Indeed, all basic building blocks are obtained by promoting Weyl spinor indices $\alpha, \dot{\alpha}$ of the Minkowski space conformal structures to $\SL(2|2)$ indices $\fa, \dfa$. For example, the simplest building block, used to absorb the conformal dimension $\Delta$ of each insertion is,
\begin{equation}
    x^2_{ij} \equiv \det( x^{\alpha \dot{\alpha}}_{ij}).
\end{equation}
The analogous structure in analytic superspace is obtained by instead taking the superdeterminant (also known as Berezinian) of the analytic superspace coordinate \eqref{eq:superspacecoord},
\begin{equation}
    \mathrm{sdet} (X_{ij}^{\fa \dfa}) = \frac{\det(x_{ij}^{\alpha \dot{\alpha}})}{\det(y_{ij}^{aa'})}+ O(\rho,\bar{\rho}).
\end{equation}
This is then used to absorb the $\gamma$ charges from each insertion.

Three-point functions of analytic superspace scalars (half-BPS supermultiplets \eqref{eq:shortdiagram}) are given by
\begin{equation}
    \langle \cO^{\gamma_1}_{\bullet} \cO^{\gamma_2}_{\bullet} \cO^{\gamma_3}_{\bullet}\rangle = \frac{C_{123}}{\mathrm{sdet} (X_{12})^{\frac12(\gamma_1+\gamma_2-\gamma_3)}\mathrm{sdet} (X_{23})^{\frac12(\gamma_2+\gamma_3-\gamma_1)}\mathrm{sdet} (X_{31})^{\frac12(\gamma_3+\gamma_1-\gamma_2)}},
\end{equation}
where $C_{123}$ is the OPE coefficient. For convenience, we define
\begin{equation}\label{eq:gij}
    g_{ij} := \mathrm{sdet}(X^{-1}_{ij}) =\frac{\det y_{ij}}{\det x_{ij}} + O(\rho,\bar{\rho}).
\end{equation}

It remains to define $\SL(2|2)\times \SL(2|2)$-valued tensors that absorb the local indices from each insertion in the correlator. At three points, there are two such covariants given by
\begin{equation}\label{eq:IijYi}
     (I_{ij})_{\dfa_i \fa_j} \sim (X_{ji}^{-1})_{\dfa_i \fa_j}, \qquad (Y_{i})_{\dfa_i \fa_i}= (X^{-1}_{ij})_{\dfa_i \fb_j} X_{jk}^{\fb_j \dfb_k} (X^{-1}_{ki})_{\dfb_k \fa_i}\sim X^{-1}_{ij} + X^{-1}_{ki},
\end{equation}
which generalise the known spinning conformal building blocks: $I_{ij}$ is the direct analogue of the conformal inversion and $Y_i$ is that of $V_i$ found in \cite{Costa:2011mg}.\footnote{A commonly used conformal building block is $H_{ij}$ which corresponds to a product of inversions $H_{ij}\propto I_{ij} I_{ji}$.}
These tensors are used to absorb all $\SL(2|2)$ indices, which must then be projected into the correct Young tableaux at each insertion. This means that indices are (anti)symmetrised according to $\ula_L$ and $\ula_R$.

General three-point functions in analytic superspace thus take the following schematic form
\begin{equation}\label{eq:threepoint}
        \langle \cO^{\gamma_1}_{\ula_L^1,\ula_R^1} \cO^{\gamma_2}_{\ula_L^2,\ula_R^2} \cO^{\gamma_3}_{\ula_L^3,\ula_R^3}\rangle = C_{123}g_{12}^{\frac12(\gamma_1+\gamma_2-\gamma_3)} g_{23}^{\frac12(\gamma_2+\gamma_3-\gamma_1)} g_{31}^{\frac12(\gamma_3+\gamma_1-\gamma_2)} \left[\prod_{ij\in S_3} I_{ij}^{a_{ij}} \prod_{k=1}^3 Y_k^{a_k} \right]_{\ula_L^k,\ula_R^k},
    \end{equation}
where $[\bullet]_{\ula}$ denotes Young symmetrisation onto $\ula$. The powers $a_{ij}, a_k$ in the tensors $I_{ij}$, $Y_k$ denote the tensor product, e.g.,
\begin{equation}
    I_{ij}^{a_{ij}} := \underbrace{I_{ij}{}_{\dfa_i\fa_j} \dots I_{ij}{}_{\dfb_i \fb_j}}_{a_{ij}},
\end{equation} 
and must satisfy
\begin{equation} \label{eq:asCondition}
    a_{ij}+a_{ik}+a_i=|\ula^i_R|, \qquad a_{ji}+a_{ki}+a_i=|\ula^i_L|\, ,
\end{equation}
with $|\ula_{L.R}^i|$ the number of boxes in the $i^{\rm th}$ Young diagram.
Throughout the paper, we make use of the following shorthand notation,
\begin{equation}\label{eq:Ysym}
     \left[Y_i\right]_{\ula} =  \big[\underbrace{Y_i{}_{\dfa \fa } \dots Y_i{}_{\dfb \fb}}_{|\ula|} \big]_{\ula}, \qquad \qquad (\ula_L=\ula_R=\ula)
\end{equation}
where the dotted and undotted indices must be symmetrised equally. 
Of particular importance to quarter-BPS correlators will be the following combination
\begin{equation}
    \left[I_{ij} I_{ji}\right]_{[1^q]} = \underbrace{I_{ij}{}_{[\dfa_i [\fa_j} \dots I_{ij}{}_{\dfb_i] \fb_j]}}_{q }\underbrace{I_{ji}{}_{[\dfa_j [\fa_i} \dots I_{ji}{}_{\dfb_j] \fb_i]}}_{q },
\end{equation}
where we recall that $\ula=[1^q]$ is a totally antisymmetrised (column) Young diagram of height $q$, characteristic of quarter-BPS operators in analytic superspace \eqref{eq:shortdiagram}.

In general, three-point functions of non-half-BPS operators admit multiple three-point structures of the form \eqref{eq:threepoint}. OPE coefficients then become degenerate and are thus labelled by an additional integer $C_{123}^{(i)}$. This is entirely analogous to three-point functions of spinning operators in CFT.

Equation \eqref{eq:threepoint} is not the most general three-point function in analytic superspace. There is the additional $\SL(2|2)$-invariant epsilon tensor $\mathcal{E}$ \cite{Heslop:2003xu} that generalises the totally antisymmetric $\epsilon$ tensor of $\SL(n)$. However, this supersymmetric epsilon tensor is an intertwiner between long representations labelled by typical Young diagrams \eqref{eq:longYD}. Thus, three-point tensor structures of short and semi-short superconformal multiplets cannot include such an invariant. 
This is important for proving non-renormalization theorems of two- and three-point functions using the Lagrangian insertion method \cite{Heslop:2001gp,Heslop:2003xu}, which we discuss in section \ref{sec:nonrenorm}.

In this paper, we will study the exchanged spectrum in four-point functions involving quarter-BPS operators.
Four-point functions can be reduced to two-point functions by means of the operator product expansion (OPE), leading to their decomposition into superconformal blocks, schematically
\begin{equation}
    \langle \cO_1 \cO_2 \cO_3 \cO_4\rangle = \sum_{i,j,k} C_{12k}^{(i)}C_{k34}^{(j)} B^{(i,j)}_k(\{\mathrm{x}_l\}),
\end{equation}
where $B^{(i,j)}_k$ is the superconformal block. The operators labelled by $k$ are known as exchanged operators. 
This is the set of operators that are at the intersection of the two OPEs defining the exchange channel. 
An operator $\cO$ appears in the OPE $\cO_1\times \cO_2$ if and only if the three-point function $\langle \cO_1 \cO_2 \cO\rangle$ admits a solution to the superconformal Ward identities.

We thus determine the exchanged spectrum by constructing three-point functions with a common exchange $\cO^{\gamma}_{\ula_L,\ula_R}$, schematically,

\begin{equation}\label{eq:gluing}
    \langle \cO_1 \cO_2 \cO^\gamma_{\ula_L,\ula_R}\rangle^{(i)} \bowtie \langle \cO^\gamma_{\ula_R,\ula_L }\cO_3 \cO_4\rangle^{(j)} \sim B^{(i,j)}_k \subseteq \langle \cO_1 \cO_2 \cO_3 \cO_4\rangle .
\end{equation}
We will denote the coordinate of the exchanged operator by $X_0$, which gets integrated over by the ``gluing'' $\bowtie$ in the full correlator. 

\subsubsection{Harmonic analyticity}\label{ssubsec:harmonic}
Once the correlator \eqref{eq:threepoint} has been set up by absorbing the charges $\gamma_i$ into the various $g_{ij}$ scalar factors and the local indices $\fa_i,\dfa_i$ into the tensors $I_{ij}$ and $Y_i$, it remains to ensure the correlator is analytic in the harmonic variables $y_{ij}$ \eqref{eq:superspacecoord}. This is shown to be equivalent to imposing the correct shortening conditions via differential Ward identities in \cite{Heslop:2001zm}.
The explicit method was first outlined in \cite{Heslop:2001gp,Heslop:2003xu} and recently used in \cite{Heslop:2025zeq}. We now summarise it using the notation of the latter.

Harmonic analyticity is first enforced on the powers of the superdeterminants $g_{ij}$ \eqref{eq:threepoint}, which must be non-negative, that is,
\begin{equation}\label{eq:positivity}
    \gamma_i+\gamma_j-\gamma_k \geq 0, \qquad i,j,k \in S_3.
\end{equation}
This is simply due to the factor $\det y_{ij}$ being in the numerator of $g_{ij}$ \eqref{eq:gij} when expanded into bosonic $2{\times}2$ blocks. A negative power would thus lead to a pole when $y_i=y_j$. This constraint selects the combinations of $\gamma_i$-charges that lead to consistent three-point functions. Indeed, when applied to both three-point functions in~\eqref{eq:gluing}, it fixes the allowed values for the charge $\gamma$ of the exchanged operator. 

To understand the selection rules for the Young diagrams $\ula^i_L,\ula^i_R$ at each insertion $i$, we must examine the pole structure of the tensorial building blocks
$I_{ij}$ and $Y_i$ \eqref{eq:IijYi}. 
They are built from inverse analytic superspace coordinates, and thus depend on $y_{ij}^{-1}$. This dependence must be cancelled by the positive powers of $\det y_{ij}$ from the scalar prefactors $g_{ij}$. For example, a structure
\begin{equation}
    g_{ij}^\gamma I_{ij}^{q_1} I_{ji}^{q_2},
\end{equation}
is analytic in the harmonic variables if $\gamma\geq q_1+q_2$, since each inverse carries a single factor of $\det y_{ij}^{-1}$. Similarly for the other basic building block,
\begin{equation}
    g_{ij}^{\gamma_1} g_{ki}^{\gamma_2} Y_i^q
\end{equation}
is analytic if $q\leq \min(\gamma_1,\gamma_2)$. This can be seen from the pole structure of $Y_i$ given in equation~\eqref{eq:IijYi}. This is enough to prove that extremal correlators of four general quarter-BPS operators do not exchange long superconformal multiplets, and derive the allowed exchanges and corresponding three-point tensor structures. This is the subject of section \ref{subsec:extremal}.

As shown in \cite{Heslop:2025zeq}, the pole(s) from the combinations of $I_{ij}, Y_i$ in a three-point structure are softened by Young projection of the indices according to the (local) Young tableaux $\ula_L^i,\ula_R^i$.

Schematically, each row in the diagram contributes a single pole in $\det y_{ij}$. The reason is that indices  are \emph{supersymmetrised}, which means the internal indices are actually antisymmetrised. And indeed any antisymmetric combination of copies of an (inverse) matrix carries a single power of the inverse determinant. Explicitly,
\begin{equation}\label{eq:softening}
    (y^{-1})^{i_1}_{[j_1}\dots (y^{-1})^{i_p}_{j_p]} \propto 
				\frac1{\det y} \epsilon_{i_1\dots i_n}y^{i_{p+1}}_{j_{p+1}}\dots y^{i_n}_{j_n}\epsilon^{j_1\dots  j_n}.
\end{equation}
This will be important in next-to-extremal correlators, studied in depth in section \ref{subsec:nextextremal}.

To determine the pole structure of a three-point tensor structure, one must examine its expansion around the locus $y_{ij}\to0$. For example, the structure
\begin{equation}\label{eq:softeningexample}
    -(I_{12})_{\dfa_1 \fa_2} (I_{21})_{\dfa_2 \fa_1}+ (Y_1)_{\dfa_1 \fa_1} (Y_2)_{\dfa_2 \fa_2}
\end{equation}
naively seems to possess a double pole at $y_{12}\to0$. However, expanding around $y_{12}\to0$ using \eqref{eq:IijYi} gives
\begin{equation} \label{eq:softeningexample2}
   \left( (y^{-1}_{12})_{\dfa_1 \fa_2} (y^{-1}_{12})_{\dfa_2 \fa_1}+ (y^{-1}_{12})_{\dfa_1 \fa_1} (y^{-1}_{12})_{\dfa_2 \fa_2} \right)(1+ O(y_{12})).
\end{equation}
The leading term is supersymmetrised in the $\SL(2|2)$ indices and therefore carries a single factor of $\det y_{12}^{-1}$. The apparent double pole has thus been softened. Such subtleties are important for determining the specific exchanges in four-point functions of general quarter-BPS operators and constructing the corresponding tensor structures \cite{Heslop:2025zeq}, but will not be needed to rule out long superconformal multiplets.

\section{Quarter-BPS correlators without a long sector}\label{sec:generalqbps}
We use the analytic superspace methods developed in section \ref{sec:superspace} to show that all extremal and next-to-extremal correlators of general quarter-BPS supermultiplets 
\begin{equation}
     \langle \cO_{[q_1,p_1,q_1]} \cO_{[q_2,p_2,q_2]} \cO_{[q_3,p_3,q_3]}\cO_{[q_4,p_4,q_4]}\rangle,
\end{equation}
do not exchange any long superconformal multiplets.
This is an important step in proving non-renormalization, since any coupling dependence can then only appear in the OPE coefficients, as will be discussed in section \ref{sec:nonrenorm}. Furthermore, we classify all exchanges in such correlators and derive the corresponding three-point tensor structures. The method is very general and can be applied to higher extremalities with a non-trivial long sector. 

\subsection{Extremal quarter-BPS correlators}\label{subsec:extremal}

We first consider the half-BPS correlator to illustrate the method, then consider the intermediate case of four quarter-BPS operators transforming in $[q_i,0,q_i]$ which very straightforwardly leads to the general $[q_i,p_i,q_i]$ case (with constraint $p_1+2q_1=p_2+p_3+p_4+2q_2+2q_3+2q_4$). We further derive all exchanged operators, which are half- and quarter-BPS.
\subsubsection{Half-BPS warmup}
Let us consider the correlator of half-BPS operators,
\begin{equation}
    \langle \cO_{p_2+p_3+p_4} \cO_{p_2} \cO_{p_3}\cO_{p_4}\rangle.
\end{equation}
In particular, we examine the intersection of two OPEs in any channel.
As explained in section \ref{subsec:Wardids},
the simplest way to achieve this is by examining the form of the two three-point functions corresponding to each OPE.
In the present case they are given by
\begin{equation}
    \begin{split}
        \langle \cO_{p_2+p_3+p_4} \cO_{p_2} \cO^\gamma_{\ula}\rangle &\propto g_{12}^{\frac12 (2p_2+p_3+p_4-\gamma)} g_{20}^{\frac12(\gamma-p_3-p_4)} g_{01}^{\frac12(\gamma+p_3+p_4)} [Y_0]_{\ula}, \\
        \langle \cO^{\gamma}_{\ula} \cO_{p_3} \cO_{p_4}\rangle &\propto g_{03}^{\frac12 (\gamma+p_3-p_4)} g_{34}^{\frac12(p_3+p_4-\gamma)} g_{40}^{\frac12(\gamma+p_4-p_3)} [Y_0]_{\ula} ,
    \end{split}
\end{equation}
where $X_0$ labels the position of the exchanged operator $\cO^\gamma_{\ula}$ and the constant of proportionality is the OPE coefficient.

Imposing positivity of the powers of all superdeterminants $g_{ij}$ \eqref{eq:positivity} we obtain
\begin{equation}
    \gamma=p_3+p_4 = p_1-p_2,
\end{equation}
which leads to the following tensor structures,
\begin{equation}
    \begin{split}
        \langle \cO_{p_2+p_3+p_4} \cO_{p_2} \cO^{p_3+p_4}_{\ula}\rangle &\propto g_{12}^{p_2}  g_{01}^{p_3+p_4} [Y_0]_{\ula}\,, \\
        \langle \cO^{p_3+p_4}_{\ula} \cO_{p_3} \cO_{p_4}\rangle &\propto g_{03}^{p_3}  g_{40}^{p_4} [Y_0]_{\ula} \,.
    \end{split}
\end{equation}
Now, recall $Y_0$ has poles in $y_{i0}$ and $y_{j0}$ \eqref{eq:IijYi}, and thus the above functions are analytic in these variables if and only if $\ula=\bullet$, the trivial representation.
Therefore, the only exchanged operator in a correlator of half-BPS operators in an extremal configuration is itself a half-BPS operator,
\begin{equation}\label{eq:pextremalex}
    \cO^{p_3+p_4}_{\bullet} \equiv \cO_{p_3+p_4}.
\end{equation}

\subsubsection{Quarter-BPS in $[q_i,0,q_i]$}
Let us consider the correlator of quarter-BPS operators,
\begin{equation}
   \langle  \cO_{[q_1,0,q_1]} \cO_{[q_2,0,q_2]} \cO_{[q_3,0,q_3]} \cO_{[q_4,0,q_4]} \rangle \equiv \langle  \cO^{2q_1}_{[1^{q_1}]}  \cO^{2q_2}_{[1^{q_2}]} \cO^{2q_3}_{[1^{q_3}]} \cO^{2q_4}_{[1^{q_4}]} \rangle, 
\end{equation}
subject to the extremality constraint $q_1=q_2+q_3+q_4$.

As in the half-BPS case above, positivity \eqref{eq:positivity} of the scalar prefactors in each three-point function \eqref{eq:gluing} yields the following charge of the exchanged operator
\begin{equation}
    \gamma=2q_3+2q_4.
\end{equation}
Said three-point functions are then given by
\begin{equation}
    \begin{split}
        \langle \cO^{2q_1}_{[1^{q_1}]}  \cO^{2q_2}_{[1^{q_2}]} \cO^{2q_3+2q_4}_{\ula}\rangle &\propto g_{12}^{2q_2}  g_{01}^{2q_3+2q_4} T^{(12)}_{[1^{q_1}],[1^{q_2}],\ula}(I_{ij},Y_i), \\
        \langle \cO_{\ula}^\gamma \cO^{2q_3}_{[1^{q_3}]} \cO^{2q_4}_{[1^{q_4}]}\rangle &\propto g_{03}^{2q_3}  g_{40}^{2q_4} T^{(34)}_{\ula,[1^{q_3}],[1^{q_4}]}(I_{ij},Y_i) ,
    \end{split}
\end{equation}
where $T^{(12)}$ and $T^{(34)}$ are tensors built from covariants of the form \eqref{eq:threepoint}.
Due to the above prefactors not containing $g_{02}$ and $g_{34}$, respectively, these tensor structures satisfy strict constraints. In particular, $T^{(12)}$ cannot depend on $I_{02},I_{20},Y_2,Y_0$ and $T^{(34)}$ on $I_{34},I_{43},Y_3,Y_4$. Moreover, as usual, all local indices must be symmetrised accordingly while being pole-free in $y_{12}$ and $y_{01}$.
We can write these constraints explicitly as follows
\begin{equation}\label{eq:3ptQQ}
    \begin{split}
        T^{(12)}_{[1^{q_1}],[1^{q_2}],\ula}&= I_{12}^{a_{12}} I_{21}^{a_{21}}  I_{01}^{a_{01}} I_{10}^{a_{10}} Y_1^{a_1}, \\
         T^{(34)}_{\ula,[1^{q_3}],[1^{q_4}]}&= I_{30}^{b_{30}} I_{03}^{b_{03}}  I_{04}^{b_{04}} I_{40}^{b_{40}} Y_0^{b_0},
    \end{split}
\end{equation}
subject to
\begin{equation}\label{eq:constraints}
    \begin{split}
        &(12):\left\{\begin{aligned}
       & a_{10}+a_{12}+a_1=a_{01}+a_{21}+a_1 = q_1\\
     &a_{12}=a_{21} = q_2 \\
     &a_{10}=a_{01}=|\ula|
    \end{aligned}\right.,\\ &(34):\left\{\begin{aligned}
       & b_{30}+b_{40}+b_0=b_{03}+b_{04}+b_0 = |\ula|\\
     &b_{30}=b_{03} = q_3 \\
     &b_{40}=b_{04}=q_4
    \end{aligned}\right..
    \end{split}
\end{equation}
Solving these (also using the extremality condition) yields
\begin{equation}\label{eq:ab}
    a_1=q_{3}+q_4-|\ula|, \quad b_0=-q_{3}-q_4+|\ula| \qquad \Rightarrow \qquad |\ula|=q_3+q_4, 
\end{equation}
since all coefficients must be non-negative. This of course implies $b_0=a_1=0$, which means that all indices in $\ula$ are contained in the $I_{0i}, I_{i0}$ within the three-point structures \eqref{eq:3ptQQ}. But these must be antisymmetrised in the $i$ (external) indices, since the external representations are quarter-BPS $\cO^{2q_i}_{[1^{q_i}]}$. This forces the indices in $\ula$ to be themselves antisymmetrised, and thus the only solution is  $\ula=[1^{q_3+q_4}]$, which corresponds to an exchange of a quarter-BPS operator
\begin{equation}\label{eq:q0qextremal}
    \cO^{2q_3+2q_4}_{[1^{q_3+q_4}]} \equiv \cO_{[q_3+q_4,0,q_3+q_4]}.
\end{equation} 
The particular form of the tensor structures is
\begin{equation}
    \begin{split}
        T^{(12)}_{[1^{q_1}],[1^{q_2}],\ula}&= [I_{12} I_{21}]_{[1^{q_2}]}  [I_{01} I_{10}]_{[1^{q_3+q_4}]}, \\
         T^{(34)}_{\ula,[1^{q_3}],[1^{q_4}]}&= [I_{03} I_{30}]_{[1^{q_3}]}  [I_{04} I_{40}]_{[1^{q_4}]}.
    \end{split}
\end{equation}

\subsubsection{Quarter-BPS in $[q_i,p_i,q_i]$}\label{ssubsec:generalextr}
We can now combine the half-BPS and quarter-BPS cases considered above to prove the most general case given by,
\begin{equation}
   \langle  \cO_{[q_1,p_1,q_1]} \cO_{[q_2,p_2,q_2]} \cO_{[q_3,p_3,q_3]} \cO_{[q_4,p_4,q_4]} \rangle \equiv \langle  \cO^{p_1+2q_1}_{[1^{q_1}]}  \cO^{p_2+2q_2}_{[1^{q_2}]} \cO^{p_3+2q_3}_{[1^{q_3}]} \cO^{p_4+2q_4}_{[1^{q_4}]} \rangle ,
\end{equation}
with $p_1+2q_1=p_2+p_3+p_4+2q_2+2q_3+2q_4$. 

Again, imposing positivity \eqref{eq:positivity} on both OPEs yields the following exchanged charge,
\begin{equation}\label{eq:gammaext}
    \gamma=p_3+p_4+2q_3+2q_4.
\end{equation}
The index counting constraints are still given by \eqref{eq:constraints} but the key difference is that $q_1$ is now given by the new extremality constraint
\begin{equation}
    q_1= \frac12 (p_2+p_3+p_4+2q_2+2q_3+2q_4-p_1).
\end{equation}
This changes the solution for $a_1$ \eqref{eq:ab} to 
\begin{equation}
    a_1=q_3+q_4 +\frac12(p_2+p_3+p_4-p_1) - |\ula|
\end{equation}
and results in $|\ula|$ being able to take multiple (integer) values in the range
\begin{equation}
    q_3+q_4\leq |\ula| \leq q_3+q_4 + \frac12 (p_2+p_3+p_4-p_1).
\end{equation}
We parametrise this by a non-negative integer $i$, $|\ula|=q_3+q_4+i$, so that the (solved) constraints for all allowed three-point tensor structures are given by
\begin{equation}
       \begin{split}
        &(12):\left\{\begin{aligned}
       & a_1 = \frac12 (p_2+p_3+p_4-p_1)-i\\
     &a_{12}=a_{21} = q_2 \\
     &a_{10}=a_{01}=q_3+q_4+i
    \end{aligned}\right.,\\ &(34):\left\{\begin{aligned}
       & b_0 =i\\
     &b_{30}=b_{03} = q_3 \\
     &b_{40}=b_{04}=q_4
    \end{aligned}\right. \qquad .
    \end{split}
\end{equation}
giving the following structures
\begin{align}
        \langle \cO^{p_1+2q_1}_{[1^{q_1}]}  \cO^{p_2+2q_2}_{[1^{q_2}]} \cO^{\tiny p_3+p_4+2q_3+2q_4}_{\ula}\rangle &\propto g_{12}^{p_2+2q_2}  g_{01}^{p_3+p_4+2q_3+2q_4} [I_{12}I_{21}]_{[1^{q_2}]} [I_{10} I_{01}]_{[1^{q_3+q_4+i}]} (Y_1)^{\frac12(p_2+p_3+p_4)-i}, \notag \\
        \langle \cO^{\tiny p_3+p_4+2q_3+2q_4}_{\ula} \cO^{p_3+2q_3}_{[1^{q_3}]} \cO^{p_4+2q_4}_{[1^{q_4}]}\rangle &\propto g_{03}^{p_3+2q_3}  g_{40}^{p_4+2q_4} [I_{30}I_{03}]_{1^{q_3}} [I_{40} I_{04}]_{[1^{q_4}]} (Y_0)^i ,\label{eq:struct3}
\end{align}
where we antisymmetrised the indices from each external insertion.
From the first we see that the exchange Young diagram $\ula$ must be completely antisymmetrised, thus giving the exchange of a quarter-BPS supermultiplet 
\begin{equation}\label{eq:qbps1}
    \cO^{p_3+p_4+2q_3+2q_4}_{[1^{q_3+q_4+i}]} \equiv \cO_{[q_3+q_4+i,p_3+p_4-2i,q_3+q_4+i]}.
\end{equation}
We have not enforced harmonic analyticity yet so the above overcounts the possible exchanges.

Having 8 parameters ($p_i$ and $q_i$) and one (extremality) constraint, there are many different configurations that lead to different ways harmonic analyticity is made manifest, thus giving different results.
However, having proved all exchanges are quarter-BPS (or half if $q_3+q_4+i=0$), it is instead much simpler to consider the intersection of the $R$-symmetry tensor products from each OPE
\begin{equation}
    [q_1,p_1,q_1]\otimes[q_2,p_2,q_2] \cap [q_3,p_3,q_3] \otimes[q_4,p_4,q_4],
\end{equation}
and directly uplift them to the corresponding quarter-BPS supermultiplets with the same $R$-symmetry.
This is because half- and quarter-BPS superprimaries are scalars whose form is fully fixed by $R$-symmetry, and thus satisfy the $R$-symmetry selection rule. 
This will not be the case for other semi-short multiplets preserving less than a quarter of the supercharges.

We will comment on the consequences of the results derived in this section for the non-renormalization properties of four-point quarter-BPS correlators in section \ref{sec:nonrenorm}, though we note here that both OPEs are extremal for all exchanged operators.

\subsection{Next-to-extremal quarter-BPS correlators}\label{subsec:nextextremal}
Like in the extremal case considered in the previous section, we build up to the analogous proof for general quarter-BPS correlators in the next-to-extremal configuration
\begin{equation}
    p_1+2q_1=p_2+p_3+p_4+2q_2+2q_3+2q_4-2.
\end{equation}
The same method becomes more complicated, as should be expected from the less constrained correlator. In particular, there is now an infinite number of representations that are exchanged. Moreover, a given exchange will in general contain multiple tensor structures (ultimately leading to multiple superconformal blocks), analogously to spinning correlators in CFT. As before, we first consider the half-BPS case, then the intermediate quarter-BPS  $[q_i,0,q_i]$ which gets easily generalised into $[q_i,p_i,q_i]$. 

\subsubsection{Half-BPS warmup}\label{subsec:half-BPSNE}
Let us first consider the following next-to-extremal correlator of four half-BPS operators, 
\begin{equation}
   \langle \cO_{p_2+p_3+p_4-2}\cO_{p_2} \cO_{p_3} \cO_{p_4} \rangle.
\end{equation}

The first novelty with respect to the extremal case is that the charge of the exchanged operator can now take on two different values
\begin{equation}
    \gamma = p_3+p_4-2+2k, \qquad k=0,1.
\end{equation}
This can be seen from the positivity constraint \eqref{eq:positivity} in the scalar prefactor, as in the extremal cases. 
We note that this multivaluedness of $\gamma$ should simply be interpreted as a statement of extremality in analytic superspace. 
Indeed, it is known that in half-BPS four-point functions, the threshold for long operators is for $k\ge 2$, which is only allowed from next-to-next-to-extremality \cite{Heslop:2001gp,Doobary:2015gia}. 

To understand what exchanged Young diagrams $\ula$ are allowed, we invoke harmonic analyticity in the (12) OPE, with three-point structure
\begin{equation}
    \langle \cO_{p_2+p_3+p_4-2}\cO_{p_2} \cO^{p_3+p_4-2+2k}_{\ula} \rangle = g_{12}^{p_2-k }g_{20}^{k} g_{01}^{p_3+p_4-2+k} [Y_0^{|\ula|}]_{\ula}.
\end{equation}
For the first choice of $k=0$, we require $\ula=\bullet$, that is, a half-BPS exchange
\begin{equation}
    \cO^{p_3+p_4-2}_{\ula=\bullet}\equiv \cO_{p_3+p_4-2}.
\end{equation}

Now, when $k=1$, non-trivial Young diagrams are allowed, subject to the constraint that $[Y_0^{|\ula|}]_{\ula}$ carries only a single pole in $\det y_{20} \in g_{20}$. Naively, it seems like $\Yboxdim{8pt}\ula=\yng(1)$ is the unique solution. However, as explained in section \ref{ssubsec:harmonic}, any single-row diagram $\ula=[\lambda]$ softens the naive higher-order pole into a simple pole. 

Therefore, the exchanged operators when $k=1$ are short and semi-short multiplets of the form
\begin{equation}\label{eq:hbps-ex}
    \begin{aligned}
        &\lambda=0: & \cO^{p_3+p_4}_{\bullet} &\equiv \cO_{p_3+p_4}  &&\text{(half-BPS)}, \\
        &\lambda=1:& \cO^{p_3+p_4}_{\yng(1)} &\equiv \cO_{[1,p_3+p_4-2,1]} &&\text{(quarter-BPS)},\\
        &\lambda \geq2:&\cO^{p_3+p_4}_{\ula=[\lambda]} &\equiv A[\lambda-2,\lambda-2]_{\Delta=p_3+p_4+\lambda-2}^{[0,p_3+p_4-2,0]} &&\text{(semi-short)},
    \end{aligned}
\end{equation}
where we used the translation of labels~\eqref{eq:reptranslation} and shortening conditions~\eqref{eq:shortdiagram}.

\subsubsection{Quarter-BPS in $[q_i,0,q_i]$}\label{ssubsec:q0qnext}
Let us consider the correlator
\begin{equation}
   \langle  \cO_{[q_1,0,q_1]} \cO_{[q_2,0,q_2]} \cO_{[q_3,0,q_3]} \cO_{[q_4,0,q_4]} \rangle \equiv \langle  \cO^{2q_1}_{[1^{q_1}]}  \cO^{2q_2}_{[1^{q_2}]} \cO^{2q_3}_{[1^{q_3}]} \cO^{2q_4}_{[1^{q_4}]} \rangle ,
\end{equation}
where $q_1=q_2+q_3+q_4-1$.

Positivity of the two OPE/three-point scalar prefactors \eqref{eq:positivity} now yields the following exchanged charges
\begin{equation}
    \gamma=2q_3+2q_4-2+2k, \qquad k=0,1.
\end{equation}
Then, the three-point functions associated with each OPE are given by
\begin{equation}\label{eq:3pt(12)(34)}
     \begin{split}
        \langle \cO^{2q_1}_{[1^{q_1}]}  \cO^{2q_2}_{[1^{q_2}]} \cO^{2q_3+2q_4-2+2k}_{\ula}\rangle &\propto g_{12}^{2q_2-k} g_{20}^{k}  g_{01}^{2q_3+2q_4-2+k} T^{(12)}_{[1^{q_1}],[1^{q_2}],\ula}(I_{ij},Y_i), \\
        \langle \cO_{\ula}^\gamma \cO^{2q_3}_{[1^{q_3}]} \cO^{2q_4}_{[1^{q_4}]}\rangle &\propto g_{03}^{2q_3-1+k} g_{34}^{1-k} g_{40}^{2q_4-1+k} T^{(34)}_{\ula,[1^{q_3}],[1^{q_4}]}(I_{ij},Y_i) .
    \end{split}
\end{equation}
Now each value of $k$ requires a longer discussion than in the half-BPS case.

\paragraph{Case 1 ($k=0$):}
Since the power of $g_{20}$ vanishes, the first structure in~\eqref{eq:3pt(12)(34)} is more constrained so we consider it first. Recall this forbids any dependence on $I_{20},I_{02},Y_0,Y_2$. Let us parametrise the structure as
\begin{equation}\label{eq:struct1}
    \langle \cO^{2q_1}_{[1^{q_1}]}  \cO^{2q_2}_{[1^{q_2}]} \cO^{2q_3+2q_4-2+k}_{\ula}\rangle \propto g_{12}^{2q_2}  g_{01}^{2q_3+2q_4-2} I_{10}^{a_{10}} I_{01}^{a_{01}} I_{12}^{a_{12}} I_{21}^{a_{21}} Y_1^{a_1}.
\end{equation}
Note this is enough to show that any possible exchange when $k=0$ is quarter-BPS, since any indices from $\ula$ can only appear in $I_{01}$ and $I_{10}$ which must be antisymmetrised.

We can go further and determine which quarter-BPS operators are exchanged. The first step is to count indices in \eqref{eq:struct1}, giving
\begin{equation}\label{eq:powers}
    a_{12}=a_{21}=q_2, \qquad a_{10}=a_{01}=|\ula|, \qquad a_1=q_{3}+q_4-1-|\ula|,
\end{equation}
which provides an upper bound for the allowed number of boxes in $\ula$.

However, by harmonic analyticity in $y_{12}$ we require
\begin{equation}
    a_{12}+a_{21}+a_1 \leq 2q_2 \quad \Rightarrow \quad a_1 = 0,
\end{equation}
giving,
\begin{equation}
    |\ula|=q_3+q_4-1.
\end{equation}

These indices must be antisymmetrised since they are contained within $I_{10}$ and $I_{01}$, which are antisymmetrised due to the quarter-BPS insertion at point 1.
Therefore, this leaves us with a single exchanged operator when $k=0$, namely the quarter-BPS multiplet given by
\begin{equation}\label{eq:k0qbpsexch}
  (k=0)\qquad \quad  \cO^{2q_3+2q_4-2}_{[1^{q_3+q_4-1}]} \equiv \cO_{[q_3+q_4-1,0,q_3+q_4-1]}.
\end{equation}

\paragraph{Case 2 ($k=1$):} Now the more constrained structure in~\eqref{eq:3pt(12)(34)} is the second, given by
\begin{equation}\label{eq:threept1}
    \langle \cO^{2q_3+2q_4}_{\ula}  \cO^{2q_3}_{[1^{q_3}]}  \cO^{2q_4}_{[1^{q_4}]} \rangle \propto g_{03}^{2q_3}  g_{40}^{2q_4} I_{30}^{b_{30}} I_{03}^{b_{03}} I_{04}^{b_{04}} I_{40}^{b_{40}} Y_0^{b_0}, 
\end{equation}
with exponents subject to
\begin{equation}\label{eq:linconst}
    \left\{\begin{aligned}
       & b_0+b_{40}+b_{30}=b_0+b_{04}+b_{03} =|\ula|\\
     &b_{30}=b_{03} = q_3 \\
     &b_{40}=b_{04}=q_4
    \end{aligned}\right. \qquad \Rightarrow \quad b_{0}=|\ula|-q_3-q_4,
\end{equation}
giving the key constraint
\begin{equation}
    |\ula|\geq q_3+q_4\,, \qquad |\ula|:=q_3+q_4+i
\end{equation}
Unlike in the $k=0$ case, the $(12)$ OPE does not give an upper bound for $|\ula|$.

Let us now examine the allowed shape of $\ula$. Using the constraints above, \eqref{eq:threept1} becomes
\begin{equation}\label{eq:34channel}
    \langle \cO^{2q_3+2q_4}_{\ula}  \cO^{2q_3}_{[1^{q_3}]}  \cO^{2q_4}_{[1^{q_4}]}\rangle \propto g_{03}^{2q_3}  g_{40}^{2q_4} [I_{30}I_{03}]_{[1^{q_3}]} [I_{04} I_{40}]_{[1^{q_4}]} Y_0^{i}, 
\end{equation}
where we included that the $I_{ij}$ tensors must be antisymmetrised due to the quarter-BPS representations at $X_3$ and $X_4$. Furthermore, note that all powers of the determinants $g_{03}$ and $g_{40}$ have been absorbed by these $I_{ij}$ poles. Therefore, this forces the remaining $|\ula|-q_3-q_4$ boxes to be symmetrised. Thus, $\ula$ must be in the following tensor product
\begin{equation}\label{eq:34lambdashape}
    \ula \in\begin{tikzpicture}[scale=0.65, baseline=(current bounding box.center)]
        \draw[thick] (0,0) rectangle (0.5,2.2);
        \draw[<->] (-0.25, 0) -- (-0.25, 2.2) node[midway,left] {$q_3$};
    \end{tikzpicture} \otimes \begin{tikzpicture}[scale=0.65, baseline=(current bounding box.center)]
        \draw[thick] (0,0) rectangle (0.5,1.8);
        \draw[<->] (-0.25, 0) -- (-0.25, 1.8) node[midway,left] {$q_4$};
    \end{tikzpicture}  \otimes \begin{tikzpicture}[scale=0.65, baseline=(align)]
        \draw[thick] (0,0) rectangle (1.8,-0.5);
        \draw[<->] (0,-0.75) -- (1.8, -0.75) node[midway,below] {$i$};
        \node (align) at (0,-0.5) {};
    \end{tikzpicture}\,.
\end{equation}
Currently there are no restrictions on the exchanged operator not being symmetric, therefore, we must leave $\ula_L\neq\ula_R$. By harmonic analyticity, the following must hold,
\begin{equation}\label{eq:heightconstraint}
    \mathrm{Height} (\ula_L) +\mathrm{Height} (\ula_R) \leq 2q_3+2q_4,
\end{equation}
where the height of a Young diagram is the height of its first column.

It remains to show that only combinations of the form $[1^\mu,\lambda]$~\eqref{eq:shortdiagram} are allowed. For this, we need to consider the (12) OPE, with corresponding three-point function given by
\begin{equation}
    \langle \cO^{2q_1}_{[1^{q_1}]}  \cO^{2q_2}_{[1^{q_2}]} \cO^{2q_3+2q_4}_{\ula}\rangle \propto g_{12}^{2q_2-1} g_{20}  g_{01}^{2q_3+2q_4-1} I_{12}^{a_{12}} I_{21}^{a_{21}} I_{20}^{a_{20}} I_{02}^{a_{02}} I_{10}^{a_{10}} I_{01}^{a_{01}}  Y_1^{a_1} Y_2^{a_2} Y_0^{a_0},
\end{equation}
subject to the following constraints
\begin{equation}\label{eq:system}
    \begin{aligned}
        a_{20} + a_{10} + a_0=a_{02} + a_{01} + a_0  &= q_3 + q_4 + i, \\
a_{02} + a_{12} + a_2=a_{20} + a_{21} + a_2  &= q_2, \\
a_{01} + a_{21} + a_1 =a_{10} + a_{12} + a_1&= q_2 + q_3 + q_4 - 1.\\
    \end{aligned}
\end{equation}
Note that since there is a single $g_{20}$, $Y_0$ must be symmetrised and there is a binary constraint on $a_{20}$ and $a_{02}$ by harmonic analyticity: they can be either $0$ or $1$. Naively, both cannot be $1$ but recall that it may be possible to construct combinations of structures with their apparent higher-order pole softened in the limit due to \eqref{eq:softening}. 

This is the first time that we will have $a_{ij}\neq a_{ji}$, which will lead to non-symmetric representations, which have different $Q$ and $\bar{Q}$ shortening conditions. In other words, they have different Young diagrams $\ula_L\neq \ula_R$ unlike all representations (exchanged and external) we have considered so far. Indeed, this  is a new feature of correlators with multiple non-half-BPS operators. 

There are four solutions depending on the chosen binary values of $a_{20},a_{02}$. Let us consider the symmetric ones first:
\begin{equation}\label{eq:symsoln}
        a_{20}=a_{02} = 0:  \qquad \left\{\begin{aligned}
            a_{01}&=a_{10}\leq q_3+q_4-1\\
            a_{12}&=a_{21}\leq q_2-1\\ 
            a_0    &= q_3 + q_4 + i - a_{10} \\
a_1    &= q_2 + q_3 + q_4 - 1 - a_{10} - a_{21} \\
a_2    &= q_2 - a_{21}
        \end{aligned}\right. ,
\end{equation}
where the bounds on the free parameters $a_{10}, a_{21}$ are due to analyticity constraints from the power of $g_{10}$ and $g_{12}$, respectively.
This solution leads to structures of the form
\begin{equation}
     g_{12}^{2q_2-1} g_{20}  g_{01}^{2q_3+2q_4-1} (I_{12} I_{21})^{a_{12}}  (I_{10}I_{01})^{a_{10}}  Y_1^{q_2+q_3+q_4-1-a_{10}-a_{12}} Y_2^{q_2-a_{12}} Y_0^{q_3+q_4+i-a_{10}}.
\end{equation}

This constrains $\ula$ to be in the following tensor product
\begin{equation}\label{eq:12lambdashape}
    \ula\  \subset \ \begin{tikzpicture}[scale=0.65, baseline=(current bounding box.center)]
        \draw[thick] (0,0) rectangle (0.5,2.2);
        \draw[<->] (-0.25, 0) -- (-0.25, 2.2) node[midway,left] {$a_{10}$};
    \end{tikzpicture}\ \ \otimes \  \begin{tikzpicture}[scale=0.65, baseline=(align)]
        \draw[thick] (0,0) rectangle (1.8,-0.5);
        \draw[<->] (0,-0.75) -- (1.8, -0.75) node[midway,below] {$|\ula|-a_{10}$};
        \node (align) at (0,-0.5) {};
    \end{tikzpicture}, \qquad   |\ula|=q_3+q_4+i,
\end{equation}
which already proves the exchange must be short since all terms in the above product are of the form $[1^{\mu},\lambda]$ \eqref{eq:shortdiagram}. 

Comparing \eqref{eq:12lambdashape} with the allowed shapes for $\ula$ coming from the $(34)$ OPE \eqref{eq:34lambdashape}, the intersections consistent with harmonic analyticity \eqref{eq:heightconstraint} can be found to be
\begin{equation}\label{eq:YD1}
    \ula= \begin{tikzpicture}[scale=0.65, baseline=(align)]
        \draw[thick] (0,0.3) rectangle (0.5,2.2);
        \draw[thick] (0.5,2.2) rectangle (2.8,1.7);
        \draw[<->] (-0.25, 0.3) -- (-0.25, 2.2) node[midway,left] { $q_3\!+\!q_4\!-\!1$};
        \draw[<->] (0.5, 2.45) -- (2.8, 2.45) node[midway,above] {$i+1$};
        \node (align) at (0,1) {};
    \end{tikzpicture},\ \begin{tikzpicture}[scale=0.65, baseline=(align)]
        \draw[thick] (0,0) rectangle (0.5,2.2);
        \draw[thick] (0.5,2.2) rectangle (2.5,1.7);
        \draw[<->] (-0.25, 0) -- (-0.25, 2.2) node[midway,left] {$q_3+q_4$};
        \draw[<->] (0.5, 2.45) -- (2.5, 2.45) node[midway,above] {$i$};
        \node (align) at (0,1) {};
    \end{tikzpicture}.
\end{equation}
The case $a_{20}=a_{02}=1$ is nearly identical and leads to the same result so we do not consider it explicitly.

Thus we turn to the non-symmetric solutions to \eqref{eq:system}, given by 
\begin{equation}\label{eq:solution}
    \begin{aligned}
        a_{20}=1,\ a_{02} &= 0:  \quad \left\{\begin{aligned}
            a_{01}&=a_{10}+1\leq q_3+q_4\\
            a_{12}&=a_{21}+1\leq q_2\\ 
            a_0    &= q_3 + q_4 + i -1- a_{10} \\
a_1    &= q_2 + q_3 + q_4 - 2 - a_{10} - a_{21} \\
a_2    &= q_2 -1- a_{21}
        \end{aligned}\right. ,
    \end{aligned}
\end{equation}
and
\begin{equation}\label{eq:solution2}
     a_{20}=0,\ a_{02} = 1: \quad \left\{\begin{aligned}
            a_{01}&=a_{10}-1\leq q_3+q_4-1\\
            a_{12}&=a_{21}-1\leq q_2-1\\ 
            a_0    &= q_3 + q_4 + i - a_{10} \\
a_1    &= q_2 + q_3 + q_4 - a_{10} - a_{21} \\
a_2    &= q_2 - a_{21}
        \end{aligned}\right. .
\end{equation}
Both solutions will be shown to be conjugate so we consider the first without loss of generality. The corresponding structure is 
\begin{equation}\label{eq:asymstructure}
     g_{12}^{2q_2-1} g_{20}  g_{01}^{2q_3+2q_4-1}I_{20} I_{12}^{a_{21}+1} I_{21}^{a_{21}}  I_{10}^{a_{10}}I_{01}^{a_{10+1}}  Y_1^{q_1-1-a_{10}-a_{21 }} Y_2^{q_2-1-a_{21}} Y_0^{|\ula|-1-a_{10}}.
\end{equation}
This structure imposes the following requirements on the exchanged Young diagrams $\ula_L,\ula_R$, which for the first time so far are allowed to be different:
\begin{equation}\label{eq:ulaLulaR}\Yboxdim{8pt}
    \ula_L \subseteq\begin{tikzpicture}[scale=0.65, baseline=(current bounding box.center)]
        \draw[thick] (0,0) rectangle (0.5,2.2);
        \draw[<->] (-0.25, 0) -- (-0.25, 2.2) node[midway,left] {$a_{10}$};
    \end{tikzpicture}\ \ \otimes \  \begin{tikzpicture}[scale=0.65, baseline=(align)]
        \draw[thick] (0,0) rectangle (1.8,-0.5);
        \draw[<->] (0,-0.75) -- (1.8, -0.75) node[midway,below] {$|\ula|-a_{10}$};
        \node (align) at (0,-0.5) {};
    \end{tikzpicture}, \qquad \ula_R \subseteq \begin{tikzpicture}[scale=0.65, baseline=(current bounding box.center)]
        \draw[thick] (0,0) rectangle (0.5,2.4);
        \draw[<->] (-0.25, 0) -- (-0.25, 2.2) node[midway,left] {$a_{10}+1$};
    \end{tikzpicture}\ \ \otimes \  \begin{tikzpicture}[scale=0.65, baseline=(align)]
        \draw[thick] (0,0) rectangle (1.6,-0.5);
        \draw[<->] (0,-0.75) -- (1.6, -0.75) node[midway,below] {$|\ula|-1-a_{10}$};
        \node (align) at (0,-0.5) {};
    \end{tikzpicture}.
\end{equation}
This already proves that all exchanges are short, since they are of the form $[1^{\mu},\lambda]$ \eqref{eq:shortdiagram}.

We can tabulate the specific exchanges by taking the intersection of \eqref{eq:ulaLulaR} with the constraint from the $(34)$ OPE \eqref{eq:34lambdashape}, given by,
\begin{equation}\label{eq:shape}
    \begin{aligned}
        \ula_L\  &  = \quad\begin{tikzpicture}[scale=0.63, baseline=(align)]
        \draw[thick] (0,0.3) rectangle (0.5,2.2);
        \draw[thick] (0.5,2.2) rectangle (2.8,1.7);
        \draw[<->] (-0.25, 0.3) -- (-0.25, 2.2) node[midway,left] { $q_3\!+\!q_4\!-\!1$};
        \draw[<->] (0.5, 2.45) -- (2.8, 2.45) node[midway,above] {$i+1$};
        \node (align) at (0,1) {};
    \end{tikzpicture}, \ \quad \begin{tikzpicture}[scale=0.63, baseline=(align)]
        \draw[thick] (0,0) rectangle (0.5,2.2);
        \draw[thick] (0.5,2.2) rectangle (2.5,1.7);
        \draw[<->] (-0.25, 0) -- (-0.25, 2.2) node[midway,left] {$q_3+q_4$};
        \draw[<->] (0.5, 2.45) -- (2.5, 2.45) node[midway,above] {$i$};
        \node (align) at (0,1) {};
    \end{tikzpicture}, \ \quad \begin{tikzpicture}[scale=0.63, baseline=(align)]
        \draw[thick] (0,-0.3) rectangle (0.5,2.2);
        \draw[thick] (0.5,2.2) rectangle (2.2,1.7);
        \draw[<->] (-0.25, -0.3) -- (-0.25, 2.2) node[midway,left] {$q_3\!+\!q_4\!+\!1$};
        \draw[<->] (0.5, 2.45) -- (2.2, 2.45) node[midway,above] {$i-1$};
        \node (align) at (0,1) {};
    \end{tikzpicture}, \\ \ula_R \  &= \quad 
    \begin{tikzpicture}[scale=0.63, baseline=(align)]
        \draw[thick] (0,0) rectangle (0.5,2.2);
        \draw[thick] (0.5,2.2) rectangle (2.5,1.7);
        \draw[<->] (-0.25, 0) -- (-0.25, 2.2) node[midway,left] {$q_3+q_4$};
        \draw[<->] (0.5, 2.45) -- (2.5, 2.45) node[midway,above] {$i$};
        \node (align) at (0,1) {};
    \end{tikzpicture}, \ \quad \begin{tikzpicture}[scale=0.63, baseline=(align)]
        \draw[thick] (0,-0.3) rectangle (0.5,2.2);
        \draw[thick] (0.5,2.2) rectangle (2.2,1.7);
        \draw[<->] (-0.25, -0.3) -- (-0.25, 2.2) node[midway,left] {$q_3\!+\!q_4\!+\!1$};
        \draw[<->] (0.5, 2.45) -- (2.2, 2.45) node[midway,above] {$i-1$};
        \node (align) at (0,1) {};
    \end{tikzpicture}\, .
    \end{aligned}
\end{equation}
The constraint from harmonic analyticity \eqref{eq:heightconstraint} rules out the third diagram for $\ula_L$ entirely, and further restricts $\ula_R=[1^{q_3+q_4+1},i-1]$ to only be paired with $\ula_L=[1^{q_3+q_4-1},i+1]$. All other combinations are allowed.

As an interesting observation, note that the asymmetric case when $i>0$ leads to superprimaries that do not appear in the OPE of two quarter-BPS superprimaries, both due to Lorentz and $R$-symmetry selection rules. However, we are considering the full superconformal OPE involving superconformal multiplets, that is, superprimaries and their superdescendants. Therefore, while the superprimary may not contribute, certain superdescendants will. Such multiplets must thus be included as genuine exchanged operators. This is another new feature of quarter-BPS OPEs compared to the half-BPS case.

The other solution in \eqref{eq:solution2} yields analogous structures for the conjugate case where $\ula_L$ and $\ula_R$ change places. When $\ula_L=\ula_R$, this simply gives additional structures for the same exchange.

All in all, we have the following exchanged operators in the $k=1$ sector ($\gamma=2q_3+2q_4$):
Quarter-BPS,
\begin{equation}\label{eq:exchq0qnext}
    \begin{aligned}
       \cO^{2q_3+2q_4}_{[1^{q_3+q_4}]} &\equiv \cO_{[q_3+q_4,0,q_3+q_4]} ,
    \end{aligned}
\end{equation}
mixed, 
\begin{equation}\label{eq:exchq0qnext2}
    \begin{aligned}
        & & \cO^{2q_3+2q_4}_{[1^{q_3+q_4}],[1^{q_3+q_4-1},1]}+c.c. &\equiv B\bar{A}[0,0]_{\Delta=2q_3+2q_4}^{[q_3+q_4,1,q_3+q_4-2]} +c.c., \\
        && \cO^{2q_3+2q_4}_{[1^{q_3+q_4-1},2],[1^{q_3+q_4+1}]}+c.c. &\equiv A \bar{B}[1,0]_{\Delta=2q_3+2q_4+\frac12}^{[q_3+q_4-2,0,q_3+q_4+1]} +c.c., 
    \end{aligned}
\end{equation}
and semi-short,
\begin{equation}\label{eq:exchq0qnext3}
    \begin{aligned}
        && \cO^{2q_3+2q_4}_{[1^{q_3+q_4},1+s]} &\equiv A[s,s]_{\Delta=2q_3+2q_4+s}^{[q_3+q_4-1,0,q_3+q_4-1]}.\\
              && \cO^{2q_3+2q_4}_{[1^{q_3+q_4-1},1+s]} &\equiv A[s,s]_{\Delta=2q_3+2q_4+s}^{[q_3+q_4-2,2,q_3+q_4-2]},\\ 
        & & \cO^{2q_3+2q_4}_{[1^{q_3+q_4-1},2+s],[1^{q_3+q_4},1+s]} &\equiv A[s+1,s]_{\Delta=2q_3+2q_4+s+\frac12}^{[q_3+q_4-2,1,q_3+q_4-1]} +c.c. ,\\ 
        & & \cO^{2q_3+2q_4}_{[1^{q_3+q_4-1},3+s],[1^{q_3+q_4+1},1+s]} &\equiv A[s+2,s]_{\Delta=2q_3+2q_4+s+1}^{[q_3+q_4-2,0,q_3+q_4]} +c.c. ,
    \end{aligned}
\end{equation}
in addition to the single exchange \eqref{eq:k0qbpsexch} from the $k=0$ sector.

\subsubsection{Four quarter-BPS in $[q_i,p_i,q_i]$}
Let us consider the most general four-point function of short representations,
\begin{equation}\label{eq:generalpqp}
   \langle  \cO_{[q_1,p_1,q_1]} \cO_{[q_2,p_2,q_2]} \cO_{[q_3,p_3,q_3]} \cO_{[q_4,p_4,q_4]} \rangle \equiv \langle  \cO^{p_1+2q_1}_{[1^{q_1}]}  \cO^{p_2+2q_2}_{[1^{q_2}]} \cO^{p_3+2q_3}_{[1^{q_3}]} \cO^{p_4+2q_4}_{[1^{q_4}]} \rangle 
\end{equation}
subject to the next-to-extremal condition
\begin{equation}
    p_1+2q_1=p_2+p_3+p_4+2q_2+2q_3+2q_4-2.
\end{equation}

As in the extremal case considered in section \ref{ssubsec:generalextr}, the general $[q_i,p_i,q_i]$ case shares many similarities with $[q_i,0,q_i]$ exercise shown above, with the replacement $2q_i\mapsto p_i+2q_i$. In particular, there are still two possible allowed values for the charge $\gamma$ of the exchanged operator now given by
\begin{equation}\label{eq:gammanext}
    \gamma=p_3+p_4+2q_3+2q_4-2+2k, \qquad k=0,1,
\end{equation}
which is a property of next-to-extremality.

The two scalar prefactors from each OPE then read
\begin{equation}\label{eq:pref}
    \begin{split}
           (12): &\qquad g_{12}^{p_2+2q_2-k}g_{20}^{k} g_{01}^{p_3+p_4+2q_3+2q_4-2+k},\\
           (34): &\qquad g_{34}^{1-k} g_{40}^{p_4+2q_4+k-1} g_{03}^{p_3+2q_3+k-1}.
    \end{split}
\end{equation}
Comparing with the scalar prefactors of the $[q_i,0,q_i]$ case \eqref{eq:3pt(12)(34)}, one should expect more allowed exchanges now since there are more powers of $g_{ij}$ to absorb poles from $I_{ij}$ and $Y_i$.
\paragraph{Case 1 ($k=0$):}

Note the tensor structure associated with the (12) OPE is given by
    \begin{equation}\label{eq:struct2}
        \langle \cO^{p_1+2q_1}_{[1^{q_1}]}\cO^{p_2+2q_2}_{[1^{q_2}]} \cO^{p_3+p_4+2q_3+2q_4-2}_{[1^{q_3+q_4-1+i}]}\rangle=
        \text{pref}\times
        [I_{12}I_{21}]_{[1^{q_2}]}[I_{10}I_{01}]_{[1^{|\ula|}]}[Y_1]_{[1^{q_1-q_2-|\ula|}]},
    \end{equation}
where the prefactor is given in \eqref{eq:pref}.
This alone proves the exchange must be quarter-BPS since all indices from $\ula$ appear in $I_{10},I_{01}$ which are antisymmetrised at point 1.
To determine the exchange, it thus suffices to take the intersection of the $R$-symmetry tensor products of each OPE and directly uplift it to the full superconformal multiplet, as in the extremal case considered in section \ref{subsec:extremal}.

\paragraph{Case 2 ($k=1$):} 
Several things remain identical to the $[q_i,0,q_i]$ case. Namely, the number of boxes in $\ula$ still is
\begin{equation}
    |\ula|=q_3+q_4+i.
\end{equation}
However, now the additional $i$ boxes are not constrained to be symmetrised like in \eqref{eq:34lambdashape}, since the  harmonic analyticity requirements are relaxed (as in the $k=0$ sector).
So far we thus have
\begin{equation}\label{eq:34lambdashape2}
    \ula \in\begin{tikzpicture}[scale=0.65, baseline=(current bounding box.center)]
        \draw[thick] (0,0) rectangle (0.5,2.2);
        \draw[<->] (-0.25, 0) -- (-0.25, 2.2) node[midway,left] {$q_3$};
    \end{tikzpicture} \otimes \begin{tikzpicture}[scale=0.65, baseline=(current bounding box.center)]
        \draw[thick] (0,0) rectangle (0.5,1.8);
        \draw[<->] (-0.25, 0) -- (-0.25, 1.8) node[midway,left] {$q_4$};
    \end{tikzpicture}  \otimes  \ \begin{tikzpicture}[scale=0.65]
        \draw[thick] (0,0) rectangle (0.5,-0.5);
    \end{tikzpicture}^{\otimes i}\,.
\end{equation}

But the linear constraints \eqref{eq:linconst} due to index counting are equivalent, and so are their solutions when kept in terms of $q_1$. Therefore, exchange Young diagrams are of the same shape, namely that of \eqref{eq:12lambdashape}, that is they are hooks $[1^\mu,\lambda]$. The only difference is that the harmonic analyticity requirement on the combined heights of $\ula_L$ and $\ula_R$ \eqref{eq:heightconstraint} is relaxed by $p_i\neq 0$, leading to more allowed representations in general.

It is not possible to derive a single analytic formula of the form \eqref{eq:exchq0qnext}-\eqref{eq:exchq0qnext3} for the general correlator \eqref{eq:generalpqp} due to the number of different configurations and orderings of the charges $p_i$ and $q_i$. Indeed, even the far simpler intersection of $R$-symmetry tensor products 
\begin{equation}\label{eq:prodinter}
    [q_1,p_1,q_1]\otimes[q_2,p_2,q_2] \cap [q_3,p_3,q_3] \otimes[q_4,p_4,q_4],
\end{equation}
does not have a closed form expression. However, the exercise presented explicitly in section \ref{ssubsec:q0qnext} can be repeated straightforwardly for other cases of interest. For example, for $q_1=p+q_2-1$ we have, in both analytic superspace and \eqref{eq:multipletnotation} \cite{Cordova:2016emh} notations,
\begin{equation}\label{eq:sum1}
    \begin{aligned}
          \cO_{[q_1,0,q_1]} \cO_{[q_2,0,q_2]}\cap \cO_{p} \cO_{p} &= \cO_{[1^{p-1}]}^{2p-2}+\sum_{s=-1} \left( \cO_{[1^{p-1},1+s]}^{2p}+\cO_{[1^{p},1+s]}^{2p} \right)\\
          &=\cO_{[p-1,0,p-1]} + \cO_{[p,0,p]} + \cO_{[p-1,2,p-1]} \\
          &\quad + \sum_{s=0} \left(A[s,s]^{[p-1,0,p-1]}_{\Delta=2p+s}+A[s,s]^{[p-2,2,p-2]}_{\Delta=2p+s}\right),
    \end{aligned}
\end{equation}
and its crossing,
\eqnal{
          \cO_{[q_1,0,q_1]}  \cO_{p}\cap \cO_{[q_2,0,q_2]}\cO_{p} &= \cO_{[1^{q_2-1}]}^{p+2q_2-2} + \sum_{s=-1} \left( \cO^{2q_2+p}_{[1^{q_2},1+s]}+\cO^{2q_2+p}_{[1^{q_2+1},1+s]} \right)\\ 
          & \quad + \sum_{s=-1} \left( \cO^{2q_2+p}_{[1^{q_2+1},1+s], [1^{q_2},2+s]} + c.c. \right)\\
          &=\cO_{[q_2-1,p,q_2-1]} + \cO_{[q_2,p,q_2]} \\ &\quad + B\bar{A}[0,0]^{[q_2+1,p-1,q_2-1]}_{\Delta=2q+p} + c.c.\\
          &\quad + \sum_{s=0} \left(A[s,s]^{[q_2-1,p,q_2-1]}_{\Delta=2q_2+p+s}+A[s,s]^{[q_2,p-2,q_2]}_{\Delta=2q_2+p+s} \right) \\ 
          & \quad + \sum_{s=0} \left(A[s,s+1]^{[q_2,p-1,q_2-1]}_{\Delta=2q_2+p+s+\frac12} +c.c. \right).
}[]
\paragraph{$R$-symmetry preserving sector from fusion rules:}
It is worth noting that, for given values of $p_i,q_i$, one can fix the sector whose superprimaries preserve the $R$-symmetry fusion rule by evaluating the tensor product \eqref{eq:prodinter} and uplifting according to the following set of rules. Recall we found that the exchanged representations in this case can be divided into isolated ($B$ shortening, half- and quarter-BPS), semi-short ($A$ shortening) or mixed ($A\bar{B}$ and $B\bar{A}$ shortening). Any term in the $R$-symmetry intersection \eqref{eq:prodinter} of the form $[q,p,r]$ with $q\neq r$ uplifts uniquely to a superconformal multiplet with mixed shortening conditions (semi-short and short) of the form
\begin{equation}
    (A\bar{B} \text{ or } B\bar{A})[0,0]^{[q,p,r]}_{\Delta=q+p+r+1} .
\end{equation}
Then, the remaining $R$-symmetry representations, which are of the form $[q,p,q]$, fall into two different categories.
First, if 
\begin{equation}
    2q+p=p_3+p_4+2q_3+2q_4,
\end{equation}
then they uplift to quarter-BPS representations,
\begin{equation}
    \cO_{[q,p,q]}.
\end{equation}
On the other hand, if
\begin{equation}
    2q+p=p_3+p_4+2q_3+2q_4-2
\end{equation}
then the exchange spectrum contains both a quarter-BPS multiplet and an infinite tower of semi-short multiplets
\begin{equation}
    \cO_{[q,p,q]}, \qquad \sum_{s=0}^{\infty}A[s,s]^{[q,p,q]}_{\Delta=p_3+p_4+2q_3+2q_4+s}.
\end{equation}
This method is thus a shortcut that avoids setting up and solving the Ward identities in analytic superspace.

\paragraph{$R$-symmetry violating sector from analytic superspace:}
The way to reconstruct the full OPE including the sector of superprimaries that violate the $R$-symmetry fusion rule (which only happens when $k=1$ \eqref{eq:gammanext}) is to first write the result in analytic superspace, combining short and semi-short representations into sums of the form \eqref{eq:sum1} by including a term with spin $s=-1$. Then, fill in the blanks in the $k=1$ sector by including all Young diagrams $\ula_L$, $\ula_R$ that are missing but whose heights are allowed by harmonic analyticity, analogously to the constraint \eqref{eq:heightconstraint}, determined by the method outlined explicitly in section \ref{ssubsec:q0qnext}.  For this, one only needs the form of the tensor structure and scalar prefactor of the $(34)$ OPE \eqref{eq:34channel} (or more generally the channel that does not include the next-to-extremal operator). This significantly reduces the amount of work needed to derive the full exchanged spectrum.

\section{Non-renormalization of quarter-BPS correlators}\label{sec:nonrenorm}

Having shown that extremal and next-to-extremal correlators of general quarter-BPS superconformal multiplets transforming in $[q_i,p_i,q_i]$ only exchange short and semi-short representations, any coupling dependence of the four-point function can only be contained in the OPE coefficients of a three-point function of the form
\begin{equation}\label{eq:3ptcoefs}
   \langle \cO_{[q_i,p_i,q_i]} \cO_{[q_j,p_j,q_j]} \cO^{\gamma}_{\ula_L,\ula_R} \rangle,
\end{equation}
where $\cO^{\gamma}_{\ula_L,\ula_R}$ is a protected operator classified in the previous section.

In extremal four-point functions (section \ref{subsec:extremal}), the relevant three-point coefficients are themselves extremal, involving only half- and quarter-BPS operators. In next-to-extremal four-point functions (section \ref{subsec:nextextremal}), one of the coefficients is extremal whereas the other is next-to-extremal, depending on the $\gamma$ sector. Therefore, it suffices to show that extremal and next-to-extremal three-point functions of short and semi-short operators are protected. This can be shown order by order by examining the structure of extremal and next-to-extremal diagrams as initiated in \cite{DHoker:2001jzy}. However, there is a stronger result in the literature.

Indeed, all three-point functions of the form \eqref{eq:3ptcoefs} for any extremality were proved to be non-renormalised in \cite{Heslop:2001gp,Heslop:2003xu,Basu:2004nt} using the Lagrangian insertion method. To see this,
let $\cO_i$, $i=1,\dots,n$, be protected operators, that is, analytic superspace fields with Young diagrams $\ula_L,\ula_R \subseteq [1^\mu,\lambda]$. Equivalently, these are operators whose shortening conditions on both sides \eqref{eq:multipletnotation} are not long \cite{Cordova:2016emh}. 
Then, the Lagrangian insertion method is as follows.
The derivative with respect to the coupling $g$ of an $n$-point function of protected operators $\cO_i$ is given by an integral of the form
\begin{equation}
    \partial_g \langle\cO_{1} \dots \cO_n \rangle \sim \int d\mu(X)  \langle\cO_{1} \dots \cO_n {\cal L}(X) \rangle  ,
\end{equation}
where ${\cal L}$ is the chiral Lagrangian at the top of the stress-tensor supermultiplet and $d\mu$ is a superspace measure including Minkowski, internal and Grassmann odd variables.
For $n=2$, the integrand
\begin{equation}
    \langle \cO_1 \cO_2 {\cal L}\rangle,
\end{equation}
can only be constructed in analytic superspace if at least one of $\cO_1,\cO_2$ is long \cite{Heslop:2003xu}, because the structure requires the supersymmetric epsilon tensor as discussed in section \ref{subsec:Wardids}. Thus the integrand is trivial for all protected operators $\cO_1,\cO_2$. The argument is identical for $n=3$; three-point structures with a Lagrangian insertion can only be constructed with this epsilon tensor, which requires at least one of the other insertions to be long. 
Therefore, two- and three-point functions of protected operators (and thus \eqref{eq:3ptcoefs}) do not depend on the coupling. 
For $n\geq 4$, the argument does not hold due to the existence of nilpotent invariants at $n+1$ points.
Semi-protected operators (long on one side and short on the other, e.g., $B{\bar L}$ in the notation \eqref{eq:multipletnotation} of \cite{Cordova:2016emh}) are not included in this argument, but also do not appear in extremal and next-to-extremal quarter-BPS four-point functions as shown in the previous section. We note that contact terms were shown to not affect the argument in \cite{Howe:1999hz,Basu:2004nt}.

In \cite{DHoker:2001jzy}, a large class of quarter-BPS three-point functions was shown to be protected at $O(g^2)$ (and arbitrary $N$) and indeed conjectured to be protected at all orders. More recently, \cite{Goncalves:2019znr, Ganchev:2025dzn} have provided further non-perturbative evidence for this using the AdS/CFT correspondence.

\section{Conclusions and outlook}\label{sec:outlook}
We showed that four-point functions of quarter-BPS operators
\begin{equation}
     \langle \cO_{[q_1,p_1,q_1]} \cO_{[q_2,p_2,q_2]} \cO_{[q_3,p_3,q_3]}\cO_{[q_4,p_4,q_4]} \rangle,
\end{equation}
are protected from quantum corrections when 
\begin{equation}
    p_1+2q_1=p_2+2q_2+p_3+2q_3+p_4+2q_4 -2k, \qquad k\in\{0,1\}.
\end{equation}
Moreover, we classified all exchanged operators (which are protected) and constructed several three-point tensor structures.

The methods considered here apply to a far wider class of correlators, both within and beyond $\cN=4$ SYM. We discuss this and other natural continuations of this work below.

First, one can consider higher extremalities. In particular, like in half-BPS correlators, we expect long superconformal multiplets to first appear at next-to-next-to-extremality. In this case, one can show that all long multiplets are contained in only one of the $\gamma$ sectors (of which there are 3 now). From this, it is straightforward to classify all long exchanges in the four-point function. This is an important precursor to a partial non-renormalization theorem for quarter-BPS correlators, generalising the seminal work of \cite{Eden:2000bk,Arutyunov:2002fh}. As shown in this work and in \cite{Hansen:2025lig}, analytic superspace is a promising formalism in which to achieve this. 

The other natural continuation is to $n$-point correlation functions of the form
\begin{equation}
    \langle \cO_{[q_1,p_1,q_1]} \cO_{[q_2,p_2,q_2]} \dots \cO_{[q_n,p_n,q_n]} \rangle,
\end{equation}
thus generalising the $n$-point half-BPS non-renormalization theorems \cite{Eden:1999kw,Eden:2000gg}. In the extremal case ($p_n+2q_n=p_1+2q_1+\dots +p_{n-1}+2q_{n-1}$) correlators were conjectured to be protected in \cite{DHoker:2001jzy}. The same methods we applied throughout the paper apply. In particular, instead of simultaneously solving two superconformal Ward identities, one has to solve $n-2$, which determine $n-3$ exchanged operators. When set up in analytic superspace, these are simply coupled linear equations on the number of indices. It is straightforward to prove in this way that extremal $n$-point functions are fully protected, and exchange only half-BPS and quarter-BPS operators, like in the four-point case. As expected, the next-to-extremal case is more involved and is left for future work.

Solving superconformal Ward identities for next-to-extremal correlators required harmonic analyticity methods developed recently in \cite{Heslop:2025zeq}. The generalisation to $d=3,6$ SCFTs is discussed in \cite{heslop:xx} and can be used to prove analogous non-renormalization theorems in 3d $\cN=8$ and 6d $\cN=(2,0)$ theories. Analytic superspace and harmonic analyticity were used in the latter case in \cite{Heslop:2004du}.

For extremal and next-to-extremal four-point functions of quarter-BPS operators,  the superconformal Ward identities force the correlator to be built from a finite set of protected superconformal blocks, so that its decomposition contains only short (semi-short) multiplets and the would-be long-multiplet contributions must vanish, reflecting the fact that these correlators are fixed by free-field theory and receive no quantum corrections. Precisely because the operator content is so constrained, these correlators can be used to read off the superconformal blocks of the protected multiplets directly, and the way these blocks organize themselves already hints at the structure of the full generic blocks for long multiplets exchanged in non-extremal correlators. It would be interesting to understand the structure of the long superconformal blocks by relaxing the extremality condition as mentioned above.  

Furthermore, having constructed the three-point functions of protected operators describing the OPEs in the relevant four-point functions, it may also be possible to directly construct the corresponding superconformal blocks by setting up and performing the shadow integral \cite{Simmons-Duffin:2012juh} in analytic superspace.

Since quarter-BPS operators are themselves multi-trace, the extremal and next-to-extremal correlators considered here give direct access to the multi-trace spectrum of protected operators. This remains valuable in particular when the extremality conditions are relaxed: the multi-trace data extracted in the protected setting provide a controlled starting point for organizing the operator content of generic correlators, and hence for extracting information about the mixing problem in general.

\section*{Acknowledgments}{We thank Francesco Aprile, Paul Heslop and Rodolfo Russo for discussions on these and related topics. Part of this project has been finalized at Amplitudes 2026, Bootstrap 2026 and Bootstrap Approaches to Scattering Amplitudes conferences, we thank all the participants for insightful discussions.
HPR is supported by a STFC studentship. GF is is supported by the US Department of Energy Office of Science under Award
Number DE-SC0015845 and by the Swiss National Science Foundation through the
project 2000-1-240037.}

\appendix
\section{Details on $\langle \CO_{[2,0,2]} \CO_2 \CO_2\CO_2 \rangle $}\label{app:explicitO02O2O2O2}
As a concrete example, in this appendix we will present a detailed computation of a next-to-extremal correlator involving one quarter-BPS and three half-BPS operators. We will first compute the four-point function explicitly in the free theory and use the existence of an underlying chiral algebra to isolate protected contributions. Then we will approach the same problem using the analytic superspace construction described in section~\ref{sec:generalqbps}. We will describe how the protected exchanged operators, fixed using analytic superspace, enter the chiral algebra decomposition of the four-point function  and how we can use that to make some predictions about the form of superblocks. 

In particular, we consider the four-point function with one insertion of the quarter-BPS operator $\CO_{[2,0,2]}$ with dimension $\Delta=4$  and three half-BPS operators $\CO_2$ in the $[0,2,0]$ representation. This is indeed an example of a next-to-extremal correlator, so we expect it to be protected, as already noticed in~\cite{Bissi:2021hjk}. We refer to~\cite{Bissi:2021hjk} for notation and further details on the implementation of the chiral algebra twist in the context of quarter-BPS operators. 

The four-point function takes the following form
\eqna{
\langle \CO_{[2,0,2]}(x_1, \mathbf{S}_1) \CO_2(x_2, y_2) \CO_2(x_3, y_3) \CO_2(x_4, y_4) \rangle= \frac{x_{24}^2}{(x_{12}^2)^3 x_{14}^2 (x_{34}^2)^2} \sum_{k=1}^3 \mathbb{T}_k \lsp \CG_{k}(z, \zb)\, ,
}[]
where $\mathbb{T}_k$ are tensors encoding  SU(4)$_R$ indices and   corresponding, respectively, to the exchange of the representations  [1,2,1], [2,0,2], [0,2,0]. By means of Wick contractions, we can compute $\CG_k$ in the free theory
\eqna{
\CG_1(z, \zb)&=\frac{\sqrt{N^2-4}}{\sqrt{3}(N^2-1)} \frac{(z\zb)^2(z\zb-z-\zb)}{(1-z)(1-\zb)}\, ,\\
\CG_2(z, \zb)&=\frac{\sqrt{N^2-4}}{\sqrt{3}(N^2-1)} \frac{(z\zb)^2(z\zb-z-\zb+2)}{(1-z)(1-\zb)}\, ,\\
\CG_3(z, \zb)&=\frac{\sqrt{N^2-4}}{\sqrt{3}(N^2-1)}\frac{z\zb  \left[z \zb (1-z)(1-\zb)+z\zb-4(1-z)(1-\zb) \right]}{(1-z)(1-\zb)}\, .
}[Gks]
Applying the chiral algebra twist $\chi$ reduces the $4d$ four-point function to a single $2d$ holomorphic function
\eqna{
\chi[\langle \CO_{[2,0,2]} \CO_2\CO_2\CO_2\rangle] \propto f(z)=-\frac{\sqrt{N^2-4}}{2\sqrt{3}(N^2-1)}\left( z^2+\frac{z}{z-1} \right)\, .
}[]
In terms of this function we can rewrite~\eqref{Gks}
\eqna{
\CG_1(z, \zb)&=\frac{2 z \zb (z f(\zb)-\zb f(z))}{z-\zb}\, , \\
\CG_2(z, \zb)&=\CA (z,\zb)\, , \\
\CG_3(z, \zb)&=\frac{4(z^2(\zb-2)f(\zb)-\zb^2(z-2)f(z))}{z-\zb}-\CA(z, \zb)\, .
}[CGAmbiguity]
We have introduced an ambiguity function $\CA(z, \zb)$, reflecting the fact that 
\eqna{
\chi \left[ \mathbb{T}_2-\mathbb{T}_3\right]=0\, .
}[]
Notice that~\eqref{CGAmbiguity} is not the only possible way of writing the solution, but the important point is that it is not possible to set $\CA=0$.\footnote{Notice that this is different from what happens in the analogous case with half-BPS operators, i.e. for $\langle \CO_4\CO_2\CO_2\CO_2\rangle$.   In this case, the chiral algebra kernel has dimension 0 and all channels can just be written as a function of $f(z, \alpha)$.}

Now we would like to understand how we can interpret this decomposition in terms of the supermultiplets we expect to exchange. First of all let us expand $f$ in SL$(2)$ blocks:
\eqna{
k_\ell^{(\Delta_{12}, \Delta_{34})}(z)&=z^\ell {}_2 F_1\left( \ell-\frac{\Delta_{12}}{2},\ell+\frac{\Delta_{34}}{2}, 2\ell, z\right)\, ,\\ 
f(z)&=\frac{\sqrt{N^2-4}}{2\sqrt{3}(N^2-1)}\left(k_1^{(2,0)}(z)+\sum_{\ell=3\text{ odd}}\frac{2 \Gamma (\ell ) \Gamma (\ell +1)}{\Gamma (2 \ell -1)} k_{\ell}^{(2,0)}(z)\right)
}[fbl]
Then we need to know that, given 
\eqna{
f(z)=b_{\ell} k_{\ell}^{(2,0)}(z)
}[fz]
when we insert this in the four-point function and expand in blocks, we get
\eqna{
\CG_1(z, \zb)|_{b_\ell}&=\begin{cases}
    0 & \ell=1\,,\\
    2^{\ell-1} g_{\tau=4, \ell-2}^{(2,0)}(z, \zb) &\ell\geq 3 \text{ odd}\, ,
\end{cases}\\
\CG_3(z, \zb)|_{b_\ell}&=\begin{cases}
    -8 g_{\tau=2, 0}^{(2,0)} & \ell=1\,,\\
    2^{\ell} g_{\tau=4, \ell-3}^{(2,0)}(z, \zb)+\frac{2^{\ell } (\ell -1) (\ell +1)}{(2 \ell -1) (2 \ell +1)}g_{\tau=4, \ell-1}^{(2,0)}(z, \zb) &\ell\geq 3 \text{ odd}\, .
\end{cases}
}[CGsdecomposition]
Finally  
\eqna{
\CA=\frac{\sqrt{N^2-4}}{2\sqrt{3}(N^2-1)} \sum_{\ell=0 \text{ even}}^\infty \frac{2^{\ell +2} \Gamma (\ell +2) \Gamma (\ell +3)}{\Gamma (2 \ell +3)} g_{\tau=4, \ell}^{(2,0)}(z, \zb)=\sum_{\ell}\CA_\ell \, g_{\tau=4, \ell}^{(2,0)}(z, \zb)\, .
}[Asdecomposition]
Notice that only one scalar twist-2 operator, which we will later identify with $\CO_2$, and twist-4 operators are exchanged. 

Now, we will use analytic superspace to identify which supermultiplets are expected to be exchanged in the OPE. The relevant three-point structures are the following
\twoseqn{
\langle\CO_{[1^2]}^4 \CO_2 \CO_{\ula}^\gamma \rangle & \propto g_{12}^{\frac{6-\gamma}{2}}g_{20}^{\frac{\gamma-2}{2}}g_{01}^{\frac{\gamma+2}{2}} T^{(12)}_{[1^2],\ula}(I_{ij}, Y_i)\, ,
}[firstEq]
{
\langle \CO_{\ula}^{\gamma}  \CO_2\CO_2\rangle &\propto g_{03}^\frac{\gamma}{2} g_{34}^{\frac{4-\gamma}{2}}g_{40}^{\frac{\gamma}{2}} [Y_0^{|\ula|}]_{\ula}\, .
}[secondEq][]
Positivity of the scalar prefactor yields
\eqna{
\gamma=2,\, 4\,,
}[gammaEx]
while imposing~\eqref{eq:asCondition} fixes
\eqna{
\begin{cases}
    a_{12}=a_{21}=a_{20}=a_{02}=a_2=0\\
    a_{01}+a_0=a_{10}+a_0=|\ula|\\
    a_{10}+a_1=a_{01}+a_1=2
\end{cases}\quad \Rightarrow \quad T^{(12),a}_{[1^2], \ula}=I_{10}^a I_{01}^a Y_1^{2-a}Y_0^{|\ula|-a}\, ,
}[]
where for simplicity we have denoted $a_{01}=a_{10}\equiv a$.

Let us first start from the second three-point function in~\eqref{secondEq}. When $\gamma=2$, the absence of poles requires $\ula=\bullet$, that is, the half-BPS exchange 
\eqna{
\CO_{\ula=\bullet}^2\equiv \CO_2\, .
}[]
Similarly to what discussed in section~\ref{subsec:half-BPSNE}, at $\gamma=4$, harmonic analyticity in the (34) OPE allows any diagram with at most two rows, and symmetry\footnote{ $Y_0\sim X_{03}^{-1}+X_{40}^{-1}$ and $Y_0|_{3\leftrightarrow 4}=-Y_0$. } under $3\leftrightarrow 4$ forces $|\ula|$ to be even. The possibility $\ula=\bullet$ is excluded because it would require $a=0$ and would produce uncompensated  poles in $y_{12}$, whereas $\ula=\square$ is excluded by $3\leftrightarrow 4$ symmetry. The remaining possibilities are 
\eqna{
\ula &= \quad  
\begin{tikzpicture}[scale=0.65, baseline=(current bounding box.center)]
        \draw[thick] (0,0) rectangle (0.5,1);
                \draw[thick] (0, 0.5)-- (0.5,0.5);
    \end{tikzpicture} \, , \qquad  
    \begin{tikzpicture}[scale=0.65, baseline=(align)]
        \draw[thick] (0,0) rectangle (3.5,0.5);
                \draw[thick] (0.5, 0)-- (0.5,0.5);
                  \draw[thick] (1, 0)-- (1,0.5);
                   \draw[<->] (1,-0.25) -- (3.5, -0.25) node[midway,below] {\footnotesize $\ell \text{ even}$};
                          \node (align) at (0,0.25) {};
    \end{tikzpicture}\, , \qquad
    \begin{tikzpicture}[scale=0.65, baseline=(align)]
        \draw[thick] (0,0) rectangle (0.5,1);
        \draw[thick] (0, 0.5)-- (0.5,0.5);
        \draw[thick] (0.5,0.5) rectangle (3.5,1);
        \draw[thick] (1, 0.5)-- (1,1);
        \draw[<->] (1,0.25) -- (3.5, 0.25) node[midway,below] {\footnotesize $\ell \text{ odd}$};
         \node (align) at (0,0.5) {};
    \end{tikzpicture} \,.
}
Let us analyse their index structure and check if they are indeed analytic. For $\ula=[1^2]$, similarly to~\eqref{eq:softening}, there is a softening of the pole and the structure 
\eqna{
T^{(12),1}_{[1^2],[1^2]}+T^{(12),2}_{[1^2],[1^2]}
}[]
is analytic, such that we can identify the corresponding exchanged operator with the quarter-BPS operator itself
\eqna{
\CO^4_{\ula=[1^2]}=\CO_{[2,0,2]}\, .
}
In the second case $\ula=[\ell+2]$, $T^{(12),2}$ vanishes identically because the indices in $I_{01}$  have to be antisymmetric at point 1 but symmetric at point 0, whereas  $T^{(12),1}_{[1^2],[\ell+2]}$ is analytic. The corresponding exchanged operator is
\eqna{
\CO_{[\ell+2]}^4=A_1[\ell, \ell ]^{[0,2,0]}_{\Delta=4+\ell} \text{ or equivalently }\CC_{[0,2,0], \ell \text{ even}}^{\frac{1}{2},\frac{1}{2}}\, .
}[]
Finally, for the last structure $\ula=[ \ell+2,1]$, both the poles in $y_{20}$ and $y_{10}$ are softened when we appropriately (anti)symmetrise indices and consider a combination of $T^{(12),1}$ and $T^{(12),2}$. Given the analyticity of this structure, it is therefore possible to exchange
\eqna{
\CO^4_{[\ell+2,1]}=A[\ell, \ell]_{\Delta=4+\ell}^{[1,0,1]} \text{ or equivalently }\CC_{[1,0,1], \ell}^{\frac{1}{4}, \frac{1}{4}}\,.
}[]
Since $[1,0,1]$ violates $R$-symmetry conservation, this is one example where only superdescendant operators will be exchanged in the OPE.

Summarizing, the only exchanged supermultiplets in $\langle \CO^4_{[1^2]}\CO^2_{\bullet} \CO_{\ula}^\gamma\rangle \bowtie \langle \CO_{\ula}^\gamma \CO^2_{\bullet} \CO^2_{\bullet} \rangle$ are
\eqna{
\CO_{\ula}^\gamma&= \lbrace \CO_{\bullet}^2, \CO_{[1^2]}^4, \CO_{[\ell+2]}^4, \CO_{[\ell+2,1]}^4 \rbrace\\
&=\lbrace {\CB^{\frac{1}{2}, \frac{1}{2}}_{[0,2,0]}, \CB_{[2,0,2]}^{\frac{1}{4}, \frac{1}{4}}, \CC_{[0,2,0], \ell \text{ even}},\CC_{[1,0,1], \ell \text{ odd}}}\rbrace\, .
}[]

Inside the four-point function in~\eqref{Gks}, both superprimaries and superdescendants of these operators can be exchanged. Considering the fact that $\CG_1$ (corresponding to the $R$-symmetry representation [1,2,1]) only exchanges twist-4 $\ell$-odd operators, $\CG_2$ ([2,0,2]) only twist-4 $\ell$-even operators and $\CG_3$ ([0,2,0]) one twist-2 scalar and twist-4 $\ell$-even operators, the only possible conformal primaries are
\begin{table}[H]
\setstretch{1.25} 
\centering
\begin{tabular}{c||c|c|c}
   op  & $\mathbb{T}_1$ & $\mathbb{T}_2$ & $\mathbb{T}_3$\\
   \hline
   \hline
$\CO_2$    &  &  & (2,0)\\
\hline
\hline
$\CO_{[2,0,2]}$ & & (4,0) &\\
$Q\bar{Q}\CO_{[2,0,2]}$ & (4,1) &  &\\
$Q^2\bar{Q}^2\CO_{[2,0,2]}$ &  & (4,2) & (4,2)\\
\hline
\hline
$\CC_{[0,2,0], \ell }$ & & & $(4, \ell)$\\
$Q \bar{Q}\CC_{[0,2,0], \ell }$ & $(4, \ell+1)$& &\\
$Q^2 \bar{Q}^2 \CC_{[0,2,0], \ell }$ & & $(4, \ell+2)$ & $3\times (4, \ell+2)$ \\
$Q^3 \bar{Q}^3 \CC_{[0,2,0], \ell }$ & $ (4, \ell+3)$  &  & \\
$Q^4 \bar{Q}^4 \CC_{[0,2,0], \ell }$ & &  & $ (4, \ell+4)$ \\
\hline \hline
$Q\bar{Q}\CC_{[1,0,1], \ell}$ & & $(4,\ell+1 )$ & $(4,\ell+1 )$\\
$Q^2\bar{Q}^2\CC_{[1,0,1], \ell}$ & $(4, \ell+2)$ &  & \\
$Q^3\bar{Q}^3\CC_{[1,0,1], \ell}$ & & $(4,\ell+3 )$ & $(4,\ell+3 )$
\end{tabular}
\end{table}
\noindent where $(\,\, ,\, )\equiv (\tau, \ell)$\,. Among these operators, the components appearing in the chiral algebra are the ones that, when decomposed in $\text{SU}(2)_R \times \text{SU}(2)_F\times \text{U(1)}_r$, are Schur operators  in the $\boldsymbol{3}$ of the SU$(2)_F$ flavor symmetry.\footnote{This is because under the breaking SU$(4)_R\to \text{SU}(2)_R \times \text{SU}(2)_F\times \text{U(1)}_r$, $\CO_2$ contains a $\CN=2$ half-BPS superprimary transforming in the $\boldsymbol{3}$ of SU(2)$_R$, $\boldsymbol{3}$ of SU$(2)_F$ and $r$-charge 0, while $\CO_{[2,0,2]}$ contains a ($r=0$)-operator in the $\boldsymbol{5}$ of SU(2)$_R$ and singlet of SU$(2)_F$. Then $(\boldsymbol{3} \otimes \boldsymbol{3}) \,  \cap \, (\boldsymbol{1} \otimes \boldsymbol{3})= \boldsymbol{3} $.} This translates in terms of SL(2) blocks as in the decomposition in~\eqref{fz} to
\eqna{
f_{\CB_{[0,2,0]}}&=\alpha_1 k_1^{(2,0)}(z)\, ,\\
f_{\CB_{[2,0,2]}}&=\alpha_2 k_3^{(2,0)}(z)\, ,\\
f_{\CC_{[0,2,0], \ell^\prime}}&=\beta_{\ell^\prime} k_{\ell^\prime+3}^{(2,0)}(z)+\tilde{\beta}_{{\ell}^\prime} k_{\ell^\prime+5}^{(2,0)}(z)\, ,\\
f_{\CC_{[1,0,1], \ell^\prime}}&=\delta_{\ell^\prime} k_{\ell^\prime+4}^{(2,0)}(z)\, .
}[]
In terms of the $b_\ell$ in~\eqref{fbl} 
\eqna{
b_1&\equiv\frac{\sqrt{N^2-4}}{2\sqrt{3}(N^2-1)}=\alpha_1\, ,\\
b_3 & \equiv b_1=\alpha_2+\beta_0\, , \\
b_{\ell\geq 5}&\equiv b_1 \frac{2\Gamma(\ell)\Gamma(\ell+1)}{\Gamma(2\ell-1)}=\beta_{\ell-3}+\tilde{\beta}_{\ell-5}+\delta_{\ell-4}\, .
}[bsCond]
Let us discuss their uplift to $4d$ one by one starting from $\CB_{[0,2,0]}$
\eqna{
f_{\CB_{[0,2,0]}} \to \mathbb{T}_3 (-8 \alpha_1 g_{\tau=2,0}^{(2,0)}(z, \zb)) 
}
plus a possible contribution from $\CA$. $\CA$ however cannot actually contribute, since it contains only twist-4 operators, which are not part of this multiplet. Moreover there is only one twist-2 block in~\eqref{CGsdecomposition}, therefore
\eqna{
\alpha_1=b_1\,,
}[]
such that ``superblock'' decomposition corresponding to the exchange of the full multiplet is
\eqna{
\CG(\CB_{[0,2,0]})=-8 b_1 \mathbb{T}_3 g_{2,0}^{(2,0)}(z, \zb)\, .
}[]
Moving to the uplift
\eqna{
f_{\CB_{[2,0,2]}} \to \alpha_2 \left[ 4\mathbb{T}_1 g_{4,1}^{(2,0)}+\mathbb{T}_3\left(8 g_{4,0}^{(2,0)}+\frac{64}{35}g_{4,2}^{(2,0)} \right)\right]+ (\mathbb{T}_2-\mathbb{T}_3)\CA_\ell g_{4, \ell}^{(2,0)}\, ,
}[]
where in this case we allowed for contributions in the chiral algebra kernel. Now notice that, from the table above, there is no component with twist four and spin 0 in [0,2,0], so this must be canceled by a term in the kernel 
\eqna{
8\alpha_2-\CA_{0}=0
}[]
which also produces a term $\mathbb{T}_2 g_{4,0}^{(2,0)}$, which was missing and correspond to the exchange of the $\CO_{[2,0,2]}$ superprimary. Using~\eqref{Asdecomposition} 
\eqna{
8\alpha_2=\CA_0=4b_1 \quad \Rightarrow \quad  \alpha_2=\frac{b_1}{2}\, .
}[alpha2]
Allowing for a further ambiguity in the coefficient of the spin-2, twist-4 superdescendant, the ``superblock'' for $\CB_{[2,0,2]}$ reads
\eqna{
\CG(\CB_{[2,0,2]})=4b_1 \mathbb{T}_2 g_{4,0}^{(2,0)}+2b_1 \mathbb{T}_1 g_{4,1}^{(2,0)}+\left[\left(\frac{32}{35}b_1-c \right)\mathbb{T}_3+c \mathbb{T}_2\right]g_{4,2}^{(2,0)}\, ,
}[]
with $c$ an unconstrained constant. 
Moving on to the semi-short multiplet $\CC_{[0,2,0], \ell \text{ even}}$, 
\eqna{
f_{\CC_{[0,2,0], \ell}} & \to \mathbb{T}_1 \left( \beta_{\ell}2^{\ell+2}g_{4,\ell+1}^{(2,0)}+\tilde{\beta}_{\ell}2^{\ell+4}g_{4, \ell+3}^{(2,0)}\right)+ \mathbb{T}_3 \Bigg[\beta_\ell 2^{\ell+3} g_{4, \ell}^{(2,0)}\\
& +2^{\ell+5} \left(\frac{\beta_{\ell}(\ell+2)(\ell+4)}{4(2\ell+5)(2\ell+7)} +\tilde{\beta}_\ell\right)g_{4, \ell+2}^{(2,0)}+\tilde{\beta}_{\ell}2^{\ell +5}\frac{ (\ell +4) (\ell +6)}{(2 \ell +9) (2 \ell +11)}g_{4, \ell+4}^{(2,0)} \Bigg]\\
&+(\mathbb{T}_2-\mathbb{T}_3)\CA^\prime_{\ell}\, g_{4,\ell+2}^{(2,0)}\, ,
}[]
and \eqref{alpha2} together with~\eqref{bsCond} implies
\eqna{
\beta_0=\frac{b_1}{2}\, .
}[]

Finally
\eqna{
f_{\CC_{[1,0,1], \ell}} & \to \mathbb{T}_1 \delta_{\ell}2^{\ell+3}g_{4,\ell+2}^{(2,0)}+ \mathbb{T}_3 \delta_\ell 2^{\ell+4}\left(g_{4,\ell+1}^{(2,0)}+\frac{(\ell +3) (\ell +5)}{(2 \ell +7) (2 \ell +9)}g_{4, \ell+3}^{(2,0)}\right)\\
&+(\mathbb{T}_2-\mathbb{T}_3)\left( \CA^{\prime\prime}_{\ell}\, g_{4,\ell+1}^{(2,0)}+\tilde{\CA}^{\prime\prime}_{\ell}\, g_{4,\ell+3}^{(2,0)} \right)\, .
}[]
Imposing consistency with~\eqref{Asdecomposition} fixes
\eqna{
\CA_{\ell}=\begin{cases}
c+\CA^\prime_0+\CA^{\prime\prime}_1 & \quad\text{for}\quad \ell=2\, ,\\
\CA^\prime_{\ell-2}+\CA^{\prime\prime}_{\ell-1}+\tilde{\CA}^{\prime\prime}_{\ell-3}& \quad\text{for}\quad \ell\geq 4 \text{ even}\, .
\end{cases}
}[AprimeCond]
The uplifts of $f$ together with the conditions~\eqref{AprimeCond} and~\eqref{bsCond} fix $\alpha_1=b_1$, $\alpha_2=\beta_0=b_1/2$ and, level by level, the sums $\beta_{\ell-3}+\tilde{\beta}_{\ell-5}+\delta_{\ell-4}$ and $\CA^\prime_{\ell-2}+\CA^{\prime\prime}_{\ell-1}+\tilde{\CA}^{\prime\prime}_{\ell-3}$, but not the individual coefficients. The relative coefficients within a given supermultiplet, such as $\tilde{\beta}_\ell/\beta_\ell$ or the kernel coefficients $c$, $\CA^\prime_\ell$, $\CA^{\prime\prime}_\ell$, $\tilde{\CA}^{\prime\prime}_\ell$, are nevertheless kinematical: since each exchanged multiplet admits a single three-point structure in analytic superspace, they are fixed by the superconformal Ward identities alone, in the same way as the coefficients of the $\langle \CO_2\CO_2\CO_2\CO_2\rangle$ superconformal blocks~\cite{Dolan:2001tt,Dolan:2004iy}. Determining them amounts to constructing the superconformal blocks for this correlator. This could be done by gluing the analytic superspace three-point structures of section~\ref{sec:generalqbps} through a shadow formalism, or, more systematically, by acting with the superconformal weight-shifting operators of~\cite{Hansen:2025lig} on the half-BPS superblocks of~\cite{Doobary:2015gia}. Once the blocks are known, \eqref{AprimeCond} and~\eqref{bsCond} determine all the OPE coefficients of the protected operators appearing in~\eqref{Gks}. We leave this to future work.
\Bibliography[ref.bib]
\end{document}